\documentclass[twocolumn]{aastex63}

\usepackage{amsmath}
\usepackage{color}
\usepackage{url}
\usepackage{ulem}
\usepackage{multirow}
\usepackage{amsmath}
\usepackage{wasysym}

\usepackage{rotating, graphicx}

\def\apjl{ApJ}%
\def\apjs{ApJS}%
\def\ao{Appl.~Opt.}%
\def\aap{A\&A}%
\newcommand\C   {\textsc{Clumpy}}

\newcommand{\reflex}{\textsc{RefleX}}

\shorttitle{Ray-tracing simulations of X-ray radiation in dusty media}
\shortauthors{Ricci et al.}

\begin{document}

\title{Ray-tracing simulations and spectral models of X-ray radiation in dusty media}

\author[0000-0001-5231-2645]{Claudio Ricci}
\affiliation{Instituto de Estudios Astrof\'{\i}sicos, Facultad de Ingenier\'{\i}a y Ciencias, Universidad Diego Portales, Avenida Ejercito Libertador 441, Santiago, Chile}
\affiliation{Kavli Institute for Astronomy and Astrophysics, Peking University, Beijing 100871, China}
\author[0000-0002-8108-9179]{St\'ephane Paltani}
\affiliation{Department of Astronomy, University of Geneva, ch. d'Ecogia 16, 1290, Versoix, Switzerland}

\correspondingauthor{Claudio Ricci \& St\'ephane Paltani}
\email{claudio.ricci@mail.udp.cl; stephane.paltani@unige.ch}

\begin{abstract}
Dust can play an important role in shaping the X-ray spectra and images of astrophysical sources. In this work we report on the implementation of dust in the ray-tracing platform \reflex. We illustrate the different effects associated to the interaction between X-ray photons and dust grains, such as dust scattering, near-edge X-ray absorption fine structures and shielding. We show how the cross-sections of the photon-gas interaction change depending on the fraction of metals in dust grains (i.e. the dust depletion factor). We compare \reflex\ simulations to the most widely-used absorption model that includes dust, and show how X-ray spectra are affected by the presence of dust in the absorbing/reprocessing medium for different geometries. We also show how \reflex\ can be used to reproduce the dust scattering halos observed in Galactic sources, and release the first torus X-ray spectral model that considers dust absorption and scattering (\textsc{RXTorusD}), to reproduce the spectra of active galactic nuclei (AGN). \textsc{RXTorusD} also considers other physical process that are not included in the most widely-used AGN torus models, such as Rayleigh scattering and scattering on molecular gas, which can lead to remarkable differences in the predicted X-ray spectra for the same set of geometrical and physical parameters.
\end{abstract}	
               

\setcounter{footnote}{0}

\section{Introduction}

X-ray emission is an ubiquitous signature of accretion, and can be observed in a wide range of astrophysical phenomena, from protoplanetary disks (e.g., \citealp{Feigelson:1999fv,Kastner:2002re}), to X-ray binaries  (e.g., \citealp{Remillard:2006pn}) and massive black holes (e.g., \citealp{Rees:1984cm,Mushotzky:1993ei}). This X-ray radiation can be strongly affected by the interstellar medium (ISM; e.g., \citealp{Ryter:1975ka,Schattenburg:1986rq,Willingale:2013fm}), and a good understanding of the effect of the ISM on the X-ray spectra of cosmic sources is crucial to both accurately constrain their intrinsic X-ray properties and to shed light on the characteristics of the ISM itself.

The ISM is composed by material in different phases: atomic, molecular and dust (e.g., \citealp{Wilms:2000vn}), and its cross-section is the sum of these three components: $\sigma_{\rm ISM}=\sigma_{\rm atomic} + \sigma_{\rm molecular} + \sigma_{\rm dust}$. Depending on its phase, the gas can have different effects on the X-ray radiation. About half of the hydrogen in the Galaxy is in atomic form, but it is not uniformly distributed \citep{Shull:1982vk}, and typically a molecular gas fraction of $f_{\rm H2}\simeq20\%$ is assumed (e.g., \citealp{Wilms:2000vn}), although in some clouds this value could be significantly higher (e.g., $f_{\rm H2}\simeq50-70\%$; \citealp{Shull:2021gk}). In the Milky Way about 1\% of the mass of the ISM is in form of dust (e.g., \citealp{Bohlin:1978bv}), a similar value is found in other galaxies with consistent metallicity (e.g., \citealp{Remy-Ruyer:2014hp}). Dust grains form from metals in the ISM, depleting them from the gas phase (e.g., \citealp{Savage:1996vh}). Depletion can be inverted when dust is destroyed, for example by blast waves from supernovae (e.g., \citealp{Jones:1994io}) or by dust sublimation (e.g., \citealp{Suganuma:2006id}). Dust participates in the different phases of the galaxy growth process. Dust grains are fundamental for gas cooling (e.g., \citealp{Schneider:2002ai,Omukai:2005sf}), which is a crucial step in the process of stellar and planet formation. Dust is also very important for the feedback process in galaxies, allowing the ISM to react to the radiation emitted by star-forming regions \citep{Murray:2005zb} or by accreting supermassive black holes (SMBHs) at relatively low values of the Eddington ratio (e.g., \citealp{Fabian:2008yd,Fabian:2009zj,Ricci:2017sj,Ricci:2022ay}). 

Dust can play an important role in the X-rays (e.g., \citealp{Wilms:2000vn,Draine:2003kb,Lee:2009gb,Costantini:2022ff}), absorbing and scattering X-ray photons. Since photoelectric absorption in the X-rays is associated to metals, having $\sim 70\%$ or more of the metals locked into the dust grains (e.g., \citealp{Wilms:2000vn}) can strongly affect the absorption process. Dust scattering is preferentially in the forward direction, which can lead to the production of diffuse halos around X-ray sources (e.g., \citealp{Rolf:1983rr}). These halos have been routinely studied over the past two decades in X-ray binaries (e.g., \citealp{Costantini:2005bx,Corrales:2015ew}), and have also been proposed as a tool to test dust models (e.g., \citealp{Smith:2006ua,Seward:2013yi}) and to measure distance of nearby galaxies (e.g., \citealp{Draine:2004zt}). The presence of dust in the ISM can also leave a clear imprint on X-ray spectra, particularly around some of the main edges. The interference pattern created by the photoelectron waves during the absorption of a X-ray photon in dust grains can in fact give rise to X-ray absorption fine structures (XAFS). These features are expected to be commonly associated to Mg, Si and O (e.g., \citealp{Costantini:2019cr,Rogantini:2019hv,Rogantini:2020lj,Psaradaki:2020tt,Zeegers:2017qe,Zeegers:2019lu}), and have been identified in the X-ray spectra of X-ray bright Galactic sources thanks to high-resolution {\it XMM-Newton} and {\it Chandra} observations (e.g., \citealp{Lee:2002bk,Lee:2010ot,Ueda:2005pp,de-Vries:2009wi,Kaastra:2009xi,Pinto:2010hn,Pinto:2013vh,Costantini:2012bi,Valencic:2013ay,Psaradaki:2022hh}). The presence of dust in the ISM could lead to significant biases when measuring the column density in the direction of X-ray sources with models that do not consider dust absorption and scattering \citep{Corrales:2016bq,Smith:2016ad}.

Most of the current X-ray models do not include dust. Notable exceptions are the widely used \textsc{tbabs} family of models \citep{Wilms:2000vn}, which however do not consider scattered radiation, and the \textsc{amol} \citep{Pinto:2010hn}, \textsc{dabs} \citep{Wilms:2000vn,Kaastra:2009xi} and \textsc{dust} \citep{Draine:2003kb} models, that are part of the \textsc{SPEX} package \citep{Kaastra:1996pf}. The \textsc{SPEX} models are based on a large effort to measure the main X-ray properties of dust in the X-ray band \citep{Costantini:2013xw}, and currently represent the state-of-the-art. However, these absorption models do not allow for complex geometries, which are often needed to reproduce all the complex features observed in the X-ray spectra of astrophysical sources. The advent of the high spectral resolution instruments on-board {\it XRISM} (Resolve, \citealp{XRISM-Science-Team:2020fo}) and {\it Athena} (X-IFU, \citealp{Barret:2016ij}) will require improved models, which include all the relevant physical processes, taking into account reprocessed radiation as well as absorbed emission, and that consider realistic geometries of the material around X-ray sources. In this work we report on the implementation of dust absorption and scattering in the ray-tracing platform \textsc{RefleX}\footnote{https://www.astro.unige.ch/reflex/} \citep{Paltani:2017fa}, which allows to create self-consistent physical models of the environment of X-ray sources including dusty gas.

\section{The new version of RefleX}\label{sec:reflex}

\textsc{RefleX} \citep{Paltani:2017fa} is a versatile ray-tracing code that allows to perform X-ray simulations over a wide range of energies (0.1\,keV--1\,MeV), considering quasi-arbitrary geometries that can be set by the user. \reflex\ allows to create complex models, that can be applied to a variety of astrophysical phenomena (e.g., \citealp{Liu:2019wv,Andonie:2022rs,McKaig:2022sj}). In the first two releases, \reflex\ included most of the physical processes relevant to radiation in the X-ray band, such as photoelectric absorption, fluorescence, Compton scattering on free electrons, Rayleigh scattering, Compton scattering on bound electrons and polarization.

\subsection{Dusty gas in \reflex}\label{sec:reflex_dust1}

The new version of the code (\reflex\ 3.0) includes the propagation of X-ray photons in a dusty medium. The parameters defining dust are however even more numerous than those defining the properties of the gas phase. Indeed dust is not only defined by the densities of the different atoms, but also by the types of molecules and crystalline structures, as well as the full distribution of grain sizes. Recent analyses \citep{Hensley2021} point towards extremely complex dust compositions in our galaxy. It would however be pointless to implement such a huge complexity in \reflex , considering that the parameter space is too large to allow us to have meaningful constraints outside of the Milky Way. For this reason, only a simple, fixed dust model is considered within \reflex , which includes nevertheless all the physical processes associated to the presence of dust.

\begin{table*}[t!]
\begin{center}
\caption{Depletion of different elements for the abundances adopted by \reflex : \textsc{aneb} \citep{Anders:1982yk}, \textsc{angr} \citep{Anders:1989ce},  \textsc{aspl} \citep{Asplund:2009ly},  \textsc{feld} \citep{Feldman:1992mc},  \textsc{grsa} \citep{Grevesse:1998zy},  \textsc{lodd} \citep{Lodders:2003vy},  \textsc{lpgp} \citep{Lodders:2009vs}, \textsc{lpgs} \citep{Lodders:2009vs}, and \textsc{wilm} \citep{Wilms:2000vn}. The depletions are reported by considering the maximum values of \textsc{dust}. The \textsc{cosmic} and \textsc{void} abundances are not reported, since dust depletion is zero for all elements. The elements not listed in the table are not depleted from the gas phase.}
\label{tab:depletion}
\begin{tabular}{lccccccccc}
\hline
\hline
\noalign{\smallskip}

Element	&  \textsc{aneb}  &  \textsc{angr}  &  \textsc{aspl}  &  \textsc{feld}  &  \textsc{grsa}  &  \textsc{lodd}  &  \textsc{lpgp}  &  \textsc{lpgs}  &  \textsc{wilm}  \\
   &	 	[\textsc{dust 1}]	&	 	[\textsc{dust 0.759}]&	 	[\textsc{dust 1}]&	 	[\textsc{dust 1}]&	 	[\textsc{dust 1}]&	 	[\textsc{dust 1}]&	 	[\textsc{dust 1}]&	 	[\textsc{dust 1}]&	 	[\textsc{dust 0.691}] \\
\noalign{\smallskip}
\hline
\noalign{\smallskip}
C		& 0.298 &  0.391 &  0.470 &  0.326 &  0.382 &  0.482 &  0.460 &  0.471 &  0.310 \\ 
O		& 0.179 &  0.167 &  0.258 &  0.152 &  0.187 &  0.241 &  0.210 &  0.216 &  0.152 \\ 
Mg		& 0.834 &  0.935 &  0.794 &  0.853 &  0.832 &  0.831 &  0.813 &  0.822 &  0.741 \\ 
Si		& 0.899 &  1.000 &  0.975 &  0.913 &  0.943 &  0.850 &  0.852 &  0.847 &  1.000 \\ 
Fe		& 1.000 &  0.759 &  1.000 &  1.000 &  1.000 &  1.000 &  1.000 &  1.000 &  0.691 \\
\noalign{\smallskip}
\hline
\noalign{\smallskip}
\end{tabular}
\end{center}
\end{table*}

The implementation of dust in \reflex\ follows \citet{Draine:2003kb}. Dust is considered as a mix of Carbon grain in graphite crystalline structure and of olivine silicate grains [Mg$_{2x}$Fe$_{2(1-x)}$SiO$_4$]. According to \citet{Draine:2003kb}, Mg and Fe are both equally represented in the dust phase of the Milky Way, thus olivine is globally in the form MgFeSiO$_4$. In addition, there is about four Carbon atoms in the dust phase for each Silicon atom \citep{Draine:2003kb}. The grain size distributions for graphite and olivine adopted in \citet{Draine:2003kb} have been derived by \cite{Weingartner:2001eb} for the Milky Way. These distributions are in the form of a power law with indices $-1.54$ for graphite and $-2.21$ for olivine starting at 3.5\,\AA\ with a modified cut-off at 0.01 and 0.164\,$\mu$m, respectively. In addition, the graphite distribution presents two additional log-normal components centered at 3.5 and 30\,\AA\ respectively. The adopted densities of graphite and silicate grains are 2.24 and 3.5 g\,cm$^{-3}$, respectively.  In \reflex , dust is governed by a single parameter (\textsc{dust}) that provides the fraction of iron in the dust phase compared to the total number of iron atoms. We will refer to this parameter as {\it dust depletion} in the following. \textsc{dust 1} means that all iron atoms are in the dust phase, so that none remains in the gas phase.  We note that compositions in \reflex\ can be set arbitrarily, and for certain abundances \textsc{dust} cannot reach one because some of the elements in the dust are fully depleted. For example, the maximum value allowed by \reflex\ for abundances from \citet{Anders:1989ce} and \citet{Wilms:2000vn} is \textsc{dust 0.759} and \textsc{dust 0.691}, respectively, for which Silicon is fully depleted. For an Iron depletion of 0.7, the depletion of Silicon is 0.9 in \citet{Wilms:2000vn}, while it is one in our dust model; this is due to a larger fraction of Fe$_2$SiO$_4$ compared to Mg$_2$SiO$_4$ in our dust with respect to the model of \citet{Wilms:2000vn}. It should be noted that \reflex\ calculates the maximum dust depletion, and will not allow the user to set a higher value of the \textsc{dust} parameter. Unless stated otherwise, throughout the paper we will use the proto-solar abundances of \citeauthor{Lodders:2009vs} (\citeyear{Lodders:2009vs}; \textsc{lpgs}). With these abundances, Carbon, Oxygen, Magnesium, and Silicon are abundant enough that all Iron atoms can be put in the dust phase (\textsc{dust 1}). This is in agreement with recent observational studies, which found that most of the iron in the ISM is found in dust (e.g., \citealp{Psaradaki:2022hh}). The depletion of the different elements found in the dust grains considered here, for different abundances and for the maximum values of \textsc{dust} allowed, are reported in Table\,\ref{tab:depletion}.

\begin{figure}
\begin{center}
\includegraphics[width=7cm]{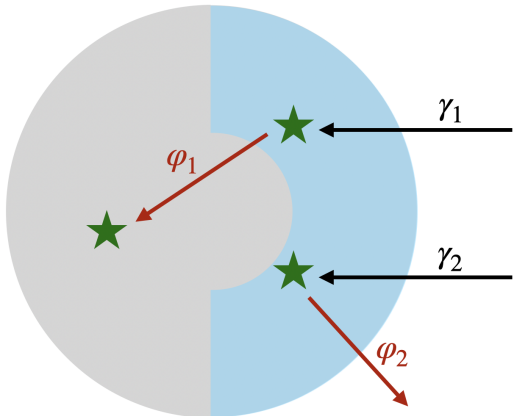}
\caption{\label{fig:grain_sketch}Sketch of the process of self-absorption of fluorescence emitted in a large grain (see \S\ref{sec:reflex_dust2}). Photons coming from the right are more likely to be absorbed in the part of the grain shown in blue, which represents schematically the mean free path for these photons. Both photons ($\gamma_1$ and $\gamma_2$) are absorbed (green stars) and fluorescence photons $\varphi_1$ and $\varphi_2$ are emitted in both cases in random directions. Photon $\varphi_1$ crosses more material than $\varphi_2$, and is more likely to be absorbed in the dust grain. Similarly, the sketch also illustrates the process of shielding (\S\ref{sec:reflex_dust2}), in which low-energy photons that hit a dust grain are absorbed in the outer part of the grain, and do not ``see'' the atoms in the center or the other side of the grain.}
\end{center}\end{figure}

\begin{figure*}
\includegraphics[width=8.8cm]{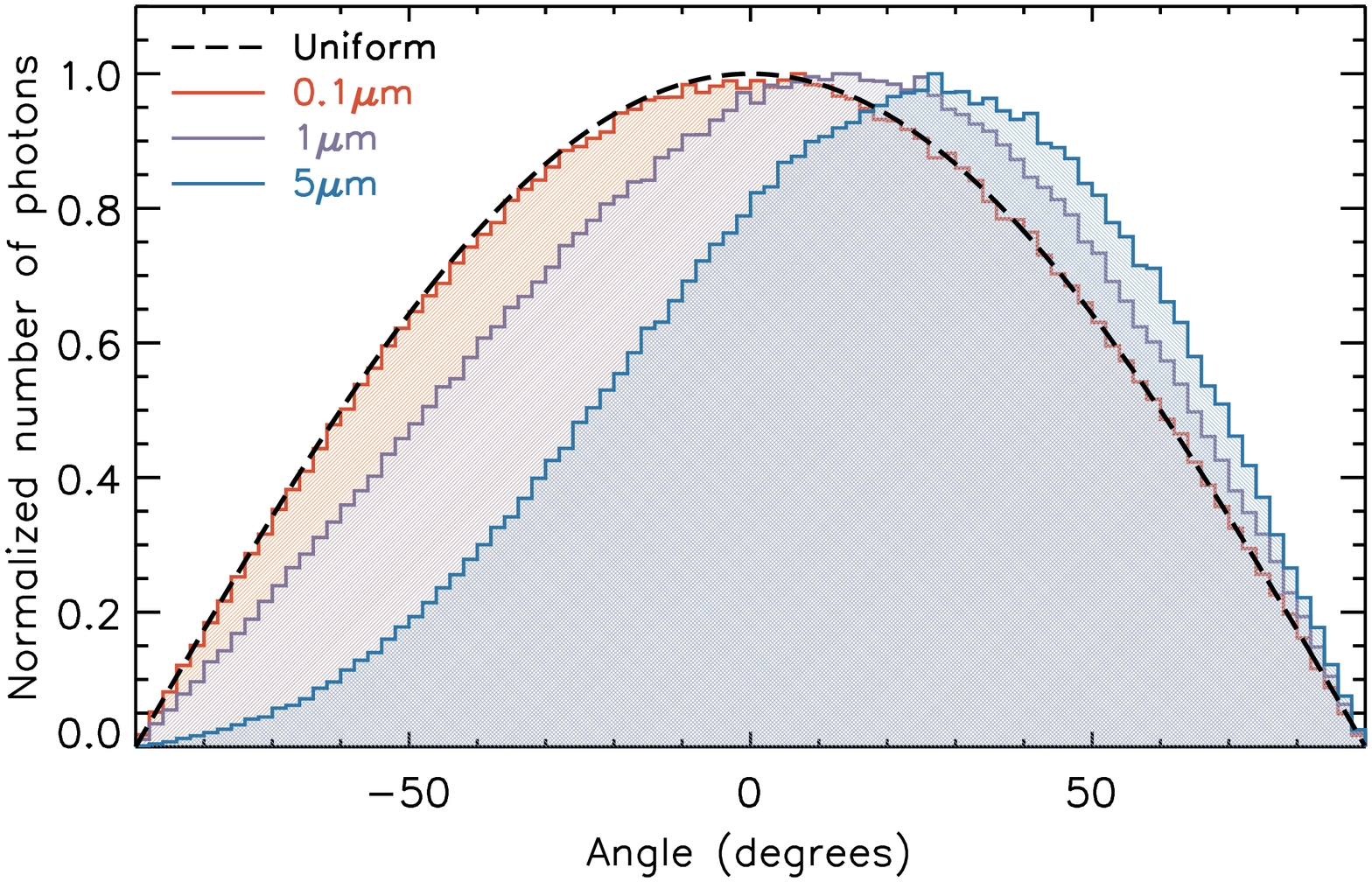}
\includegraphics[width=8.8cm]{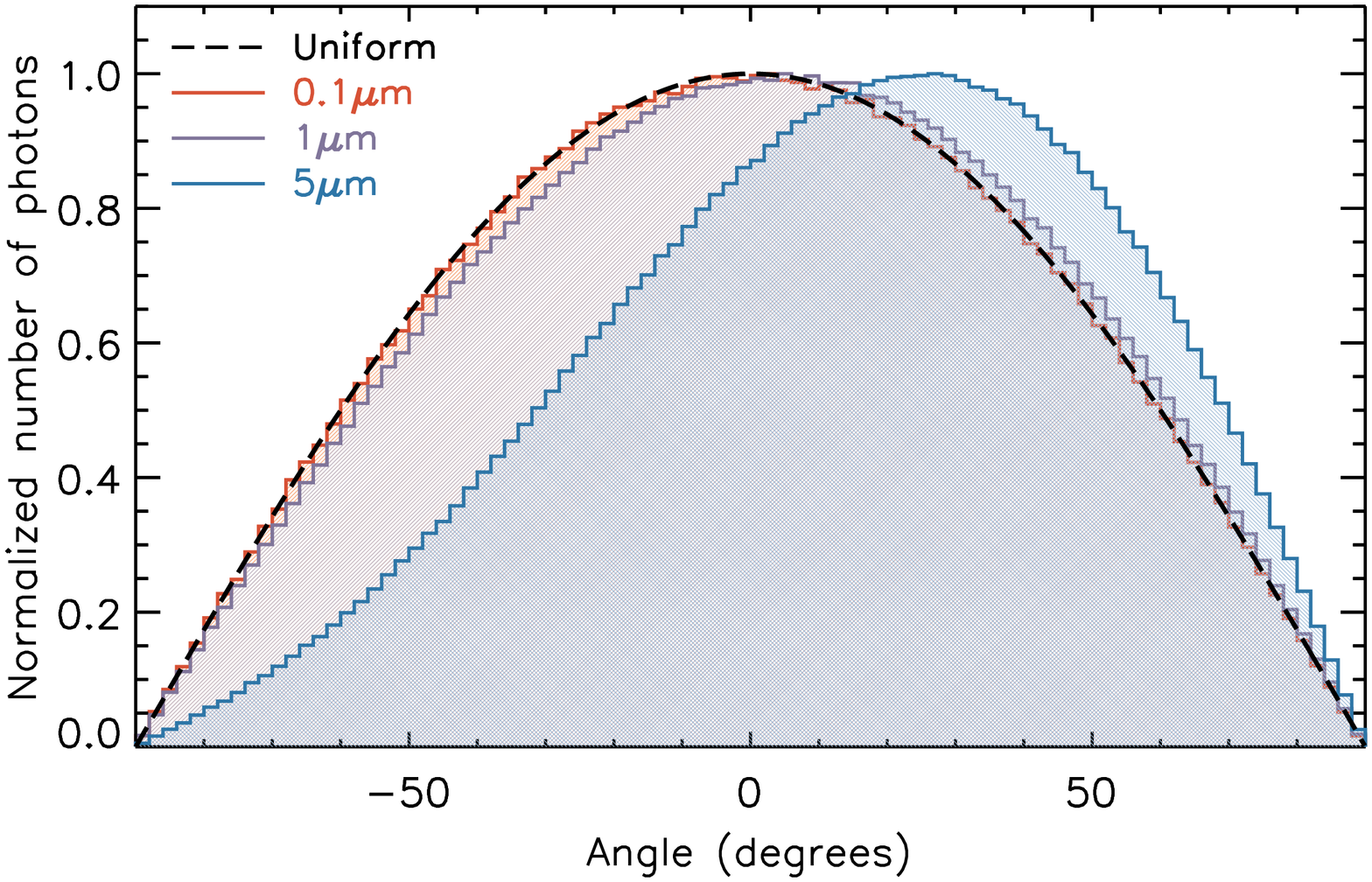}
\caption{Angular distributions of fluorescence photons emitted by Carbon (left panel) and Silicon (right panel) in a dust grain illuminated by 0.3\,keV (C) or 2\,keV (Si) photons for different grain radii: 0.1$\mu$m (red line), 1$\mu$m (purple line) and 5$\mu$m (blue line). The black line shows the angular distribution for an isotropic source of radiation. An angle of $-90^{\circ}$ means that the photon direction is opposite to that of the incoming photons. The histograms are normalised to peak at unity.}\label{fig:grain_angular}
\end{figure*}

\begin{figure*}
\centering
\includegraphics[width=0.49\textwidth]{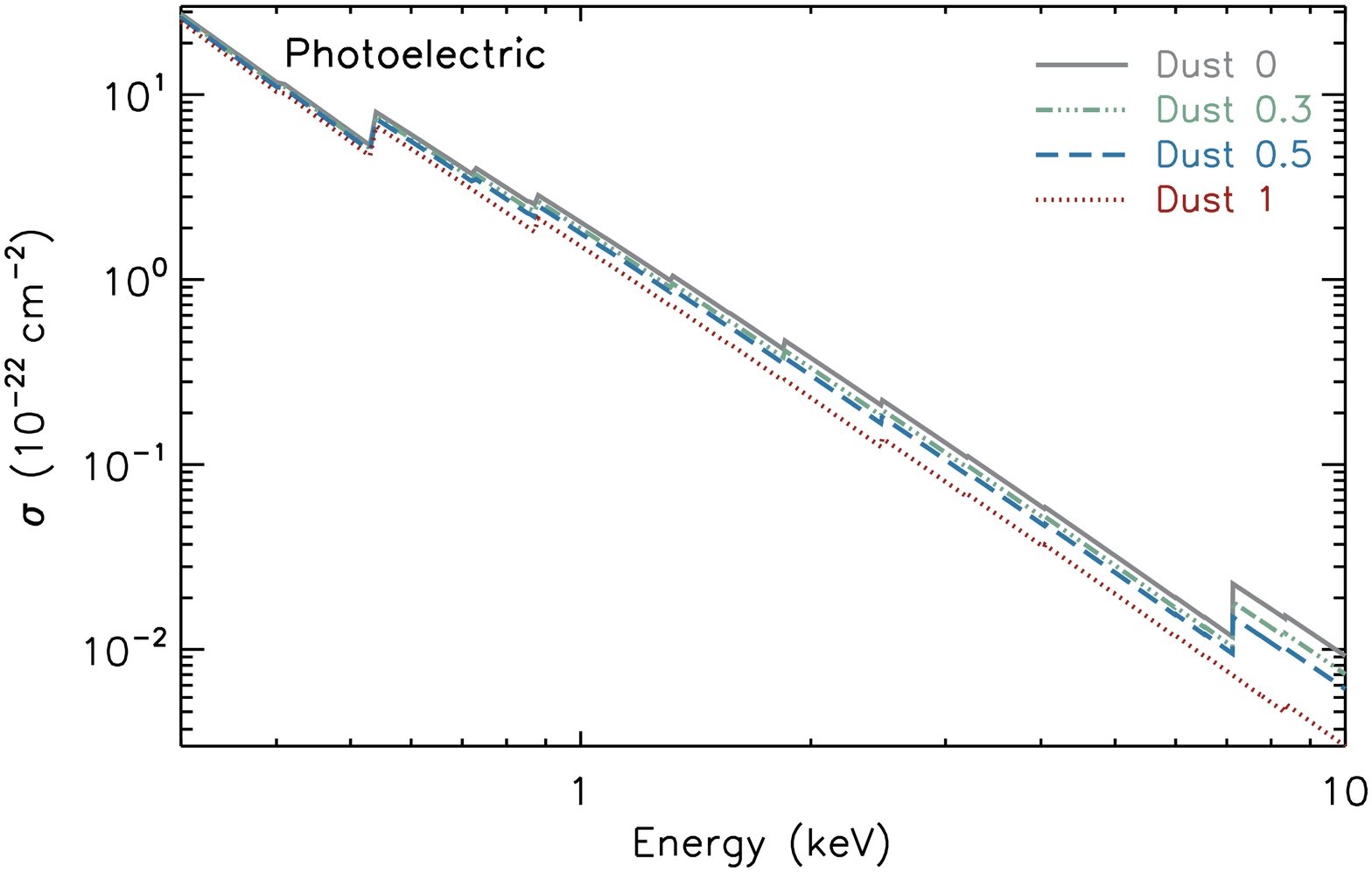} 
\includegraphics[width=0.49\textwidth]{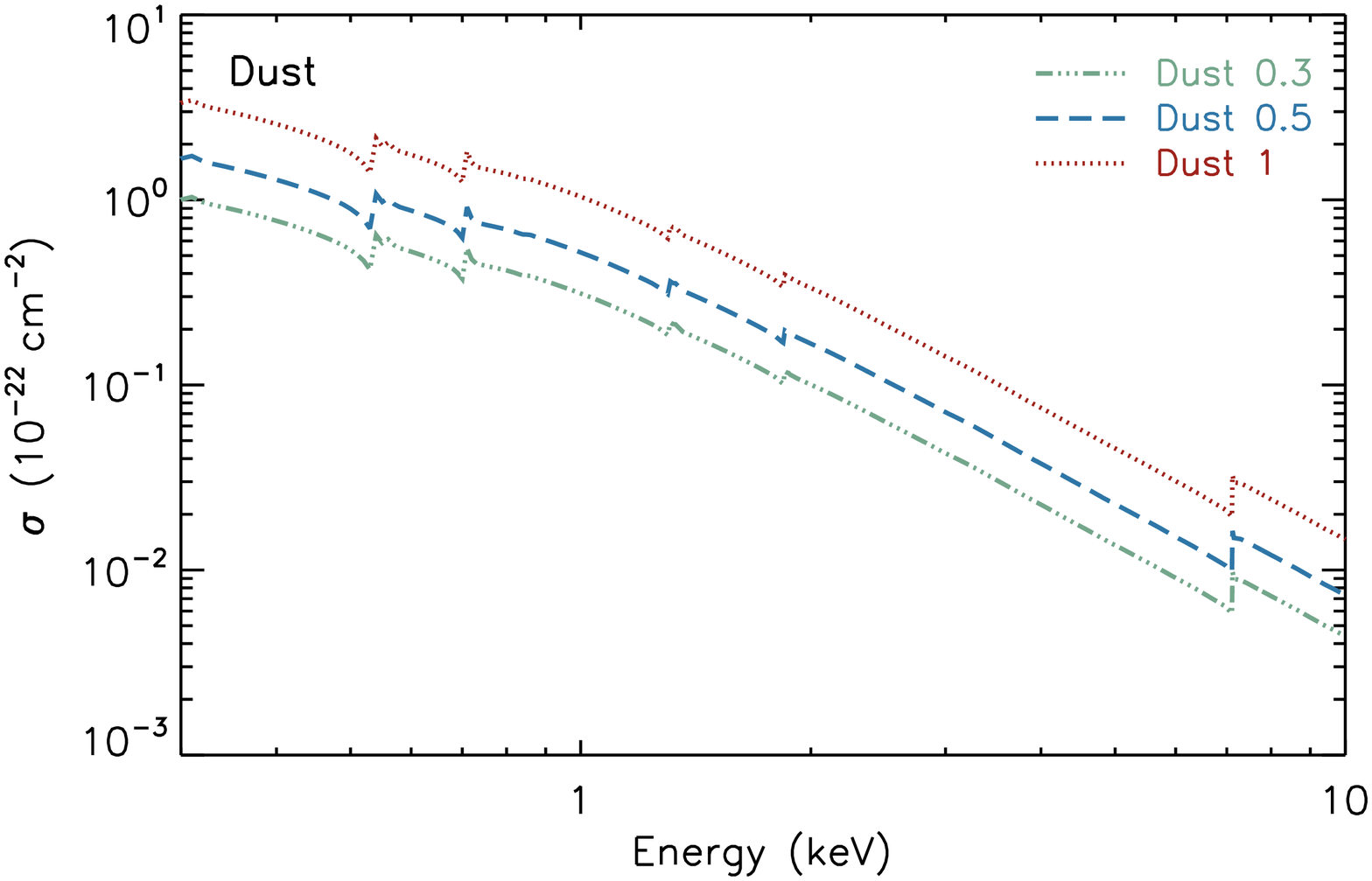} 
\par\smallskip
\includegraphics[width=0.49\textwidth]{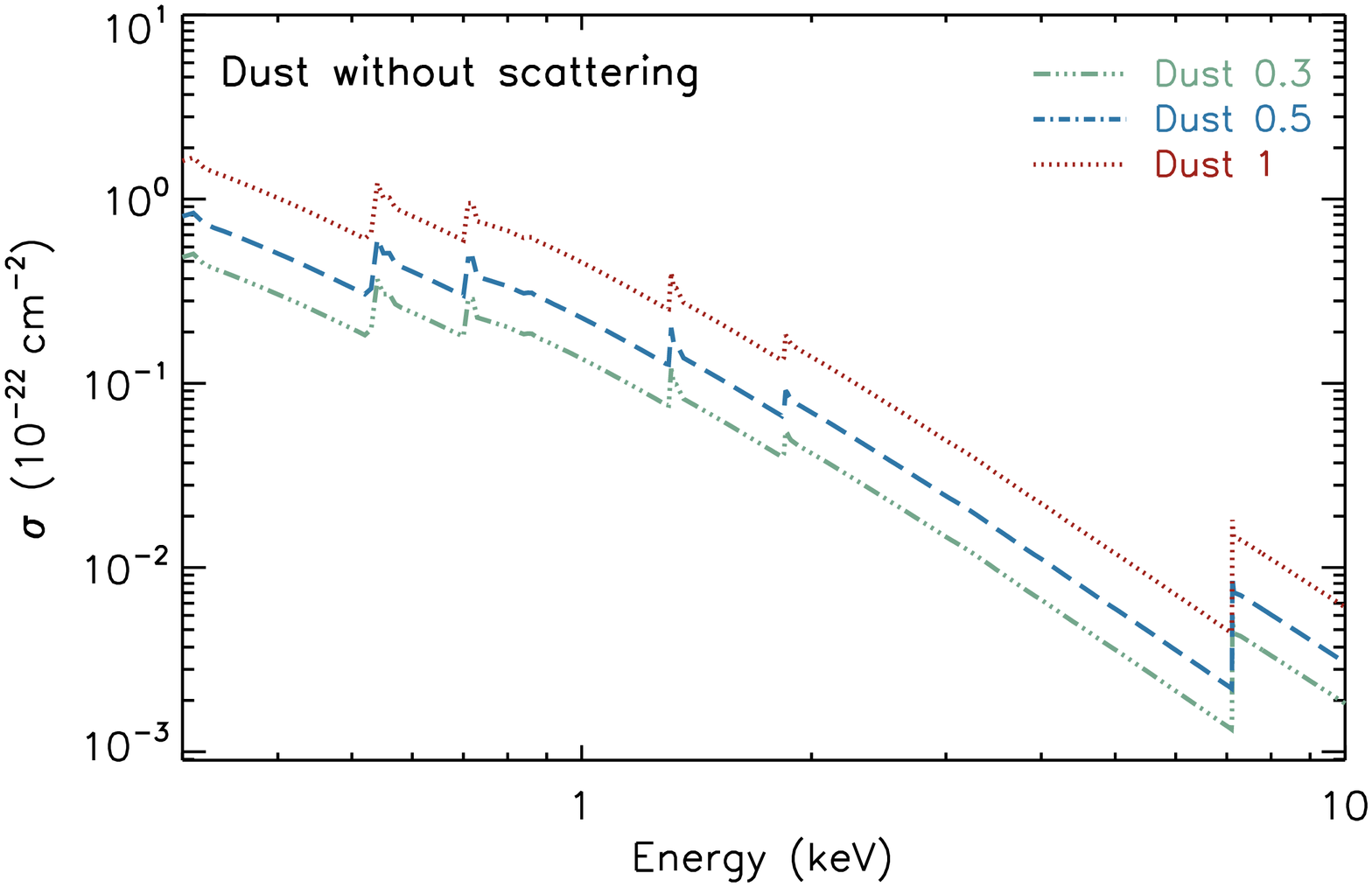} 
\includegraphics[width=0.49\textwidth]{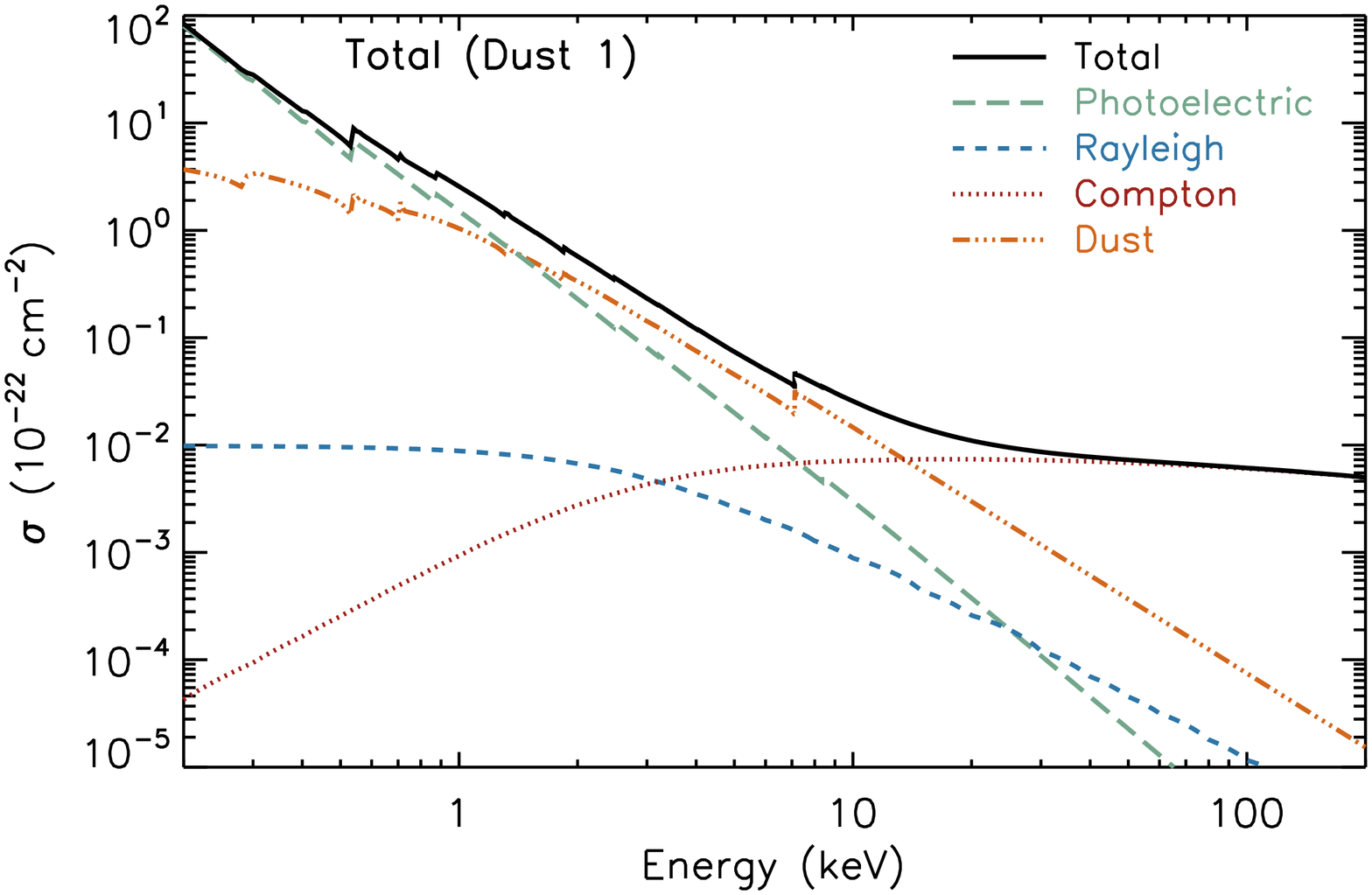} 
\caption{Different cross-sections of gas used in \reflex . {\it Top left panel:} photoelectric cross-section for gas with no dust (continuous grey line), and dust depletion 0.3 (green dot-dot-dashed line), 0.5 (blue dashed line) and 1 (red dotted line). {\it Top right panel:} same as the top left panel for the dust cross section. {\it Bottom left panel:} same as the top right panel excluding dust scattering. {\it Bottom right panel:} Total cross section of dusty gas for \textsc{dust 1} (black solid line), together with the contribution of photoelectric absorption (long-dashed green line), Rayleigh scattering (dashed cyan line), Compton scattering (dotted red line), and dust (dot-dot-dashed orange line). The proto-solar abundances of \citet{Lodders:2009vs} were considered to calculate the cross-sections. The spectral resolution was set to 10\,eV.}
\label{fig:crosssections}
\end{figure*}

\subsection{New physical processes included}\label{sec:reflex_dust2}

Three physical processes are specific to the presence of dust in the X-rays, and are implemented in \reflex :
\begin{itemize}
	\item \textbf{Scattering on dust grains.} While Rayleigh scattering can be considered as scattering on particles of sizes of the same order of magnitude as, or smaller than, the wavelength of the incoming photon, dust particles are significantly larger than the photon wavelength, but not so large that scattering becomes normal optical reflection. Scattering on dust is described by Mie \citep{Wiscombe:1980aa} and anomalous diffraction (ADT; \citealp{van-de-Hulst:1957fg}) theory. For dust grains of radii $a$, Mie theory is used for $2\pi a/\lambda<2\times10^4$, while ADT is considered for $2\pi a/\lambda>2\times10^4$. Dust scattering is elastic, and mostly forward, with the median scattering angle decreasing as the photon energy increases (Eq.\,\ref{eq:scatteringangle}). The cross-sections and differential cross-section used here are taken from \citet{Draine:2003kb}.
	\item \textbf{NEXAFS} (Near-Edge X-ray Absorption Fine Structure). The crystalline arrangement of the atoms in dust grains affects the photo-ionization absorption cross-sections near the X-ray edges very close to the transition energy. The precise cross-sections can be computed using the GGADT code \citep{Hoffman:2016po} for any dust composition, shape and size, and could constitute a diagnostic of the precise dust content using very sensitive X-ray spectrometers, like the X-IFU instrument \citep{Barret2018} on the future X-ray mission Athena \citep{Barcons2017}. The photo-ionization absorption cross-sections in dust used in \reflex\ are also obtained from \citet{Draine:2003kb}, which includes the effect of NEXAFS.
	\item \textbf{Shielding}. Self-shielding is another process that affects the photo-ionization absorption cross-sections of dust, and is the effect by which some atoms are hidden from X-ray radiation. This is due to the fact that atoms in dust are very strongly clustered in grains. Therefore, low-energy photons that hit a dust grain will be absorbed in the front part of the grain, and will never ``see'' the atoms on the other side of the grain (see Fig.\,\ref{fig:grain_sketch}). The effect of shielding is to decrease the photo-ionization absorption cross-sections at low energies ($\lesssim 3$\,keV); at higher energies ($\gtrsim 3$\,keV), for the dust grain sizes considered here, shielding is minimal (e.g., \citealp{Wilms:2000vn}). Shielding is implemented in the cross-sections of \citet{Draine:2003kb} that are used in \reflex .
\end{itemize}

Dust could also affect the process of fluorescence. Fluorescence from dusty gas could be expected in different astrophysical environments around bright X-ray sources, such as for example a torus around accreting supermassive black holes (e.g., \citealp{Antonucci93,Netzer:2013ix,Ramos-Almeida:2017hw}).
When a photon is absorbed in an optically-thick grain, absorption will not occur uniformly in the grain, but it will happen more likely close to the surface of the grain. Fluorescent photons are emitted isotropically. However, when a fluorescence photon is created inside a dust grain, the path length through the grain will depend on the direction of the reemitted photon, and there would some directions where the fluorescence photon has more chance of being reabsorbed immediately, which would affect the observed angular distribution of fluorescent photons. This is illustrated in Fig.~\ref{fig:grain_sketch}. The effect depends on the size of the grain and the energy of the photons, and will be negligible for the smallest grains. In order to establish the importance of this effect, we simulated using \reflex\ a single grain that we illuminate with X-rays, and checked the direction of outgoing fluorescence photons. We performed a number of simulations of single Carbon and Silicon grains with different photon energies and grain sizes. Figure~\ref{fig:grain_angular} shows the angular distributions of fluorescence photons (mostly C\,K$\alpha$ and Si\,K$\alpha$) in the case of grains illuminated by 0.3\,keV (C) or 2\,keV (Si) photons for spherical grains with radius 0.1\,$\mu$m, 1\,$\mu$m, and 5\,$\mu$m, respectively. Here we focus primarily on dust grain in the ISM and, at sizes of the order of the maximum grain size in our adopted distribution, no deviation from an isotropic distribution is observed; however, for larger grains, the vast majority of photons are emitted backwards, as expected. We conclude that, for the ISM grains considered here, this effect is insignificant, and thus it is not implemented in \reflex. We point out that this result is not unexpected, as the fluorescence photons are emitted below their respective edge energies, and encounter therefore significantly lower photo-electric cross-sections. We discuss in detail in \S\ref{sec:fluorescent} the relation between dust depletion and the production of fluorescent lines.

\begin{figure*}[t]
\centering
\includegraphics[width=0.49\textwidth]{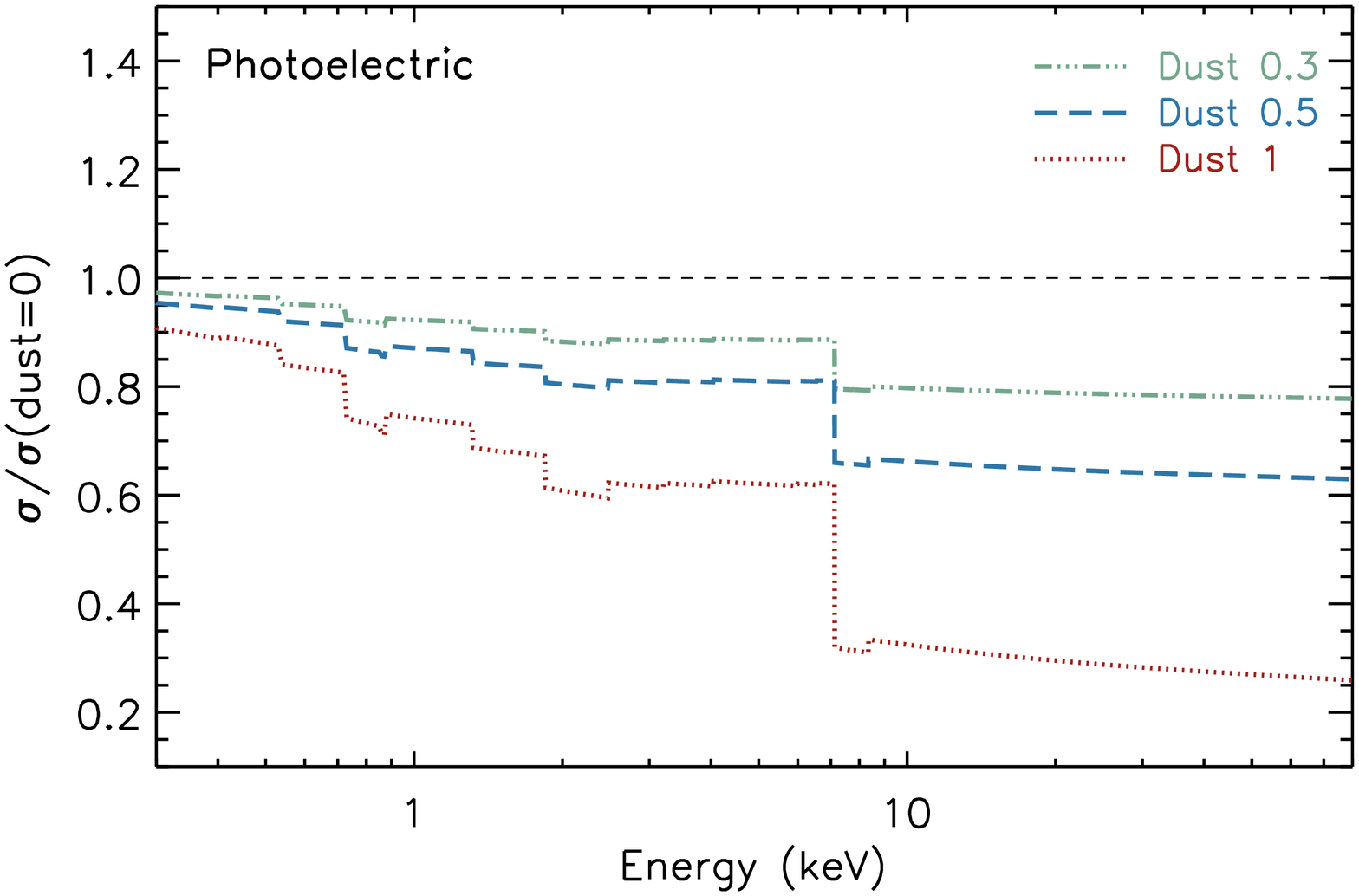} 
\includegraphics[width=0.49\textwidth]{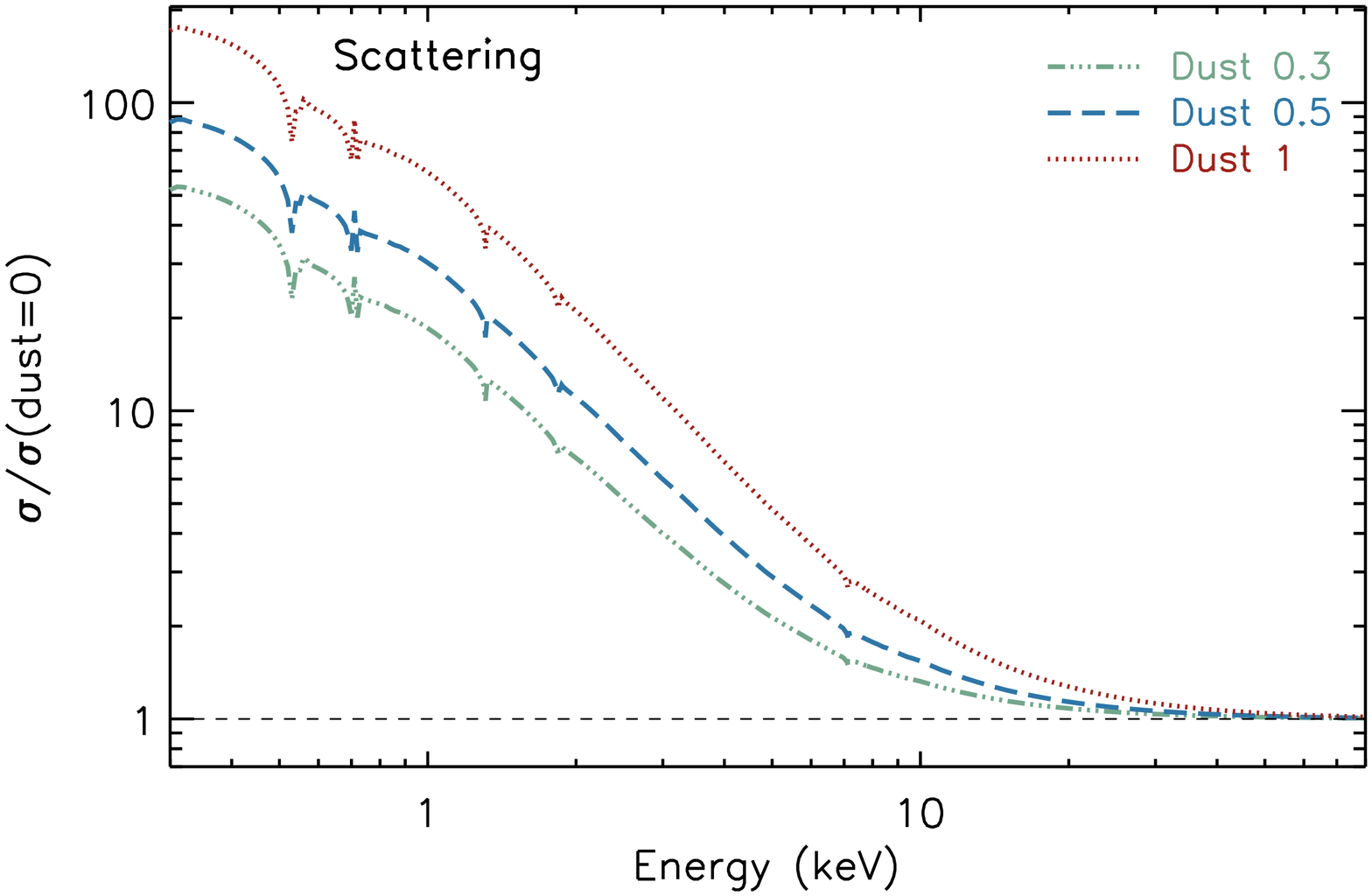} 
\includegraphics[width=0.49\textwidth]{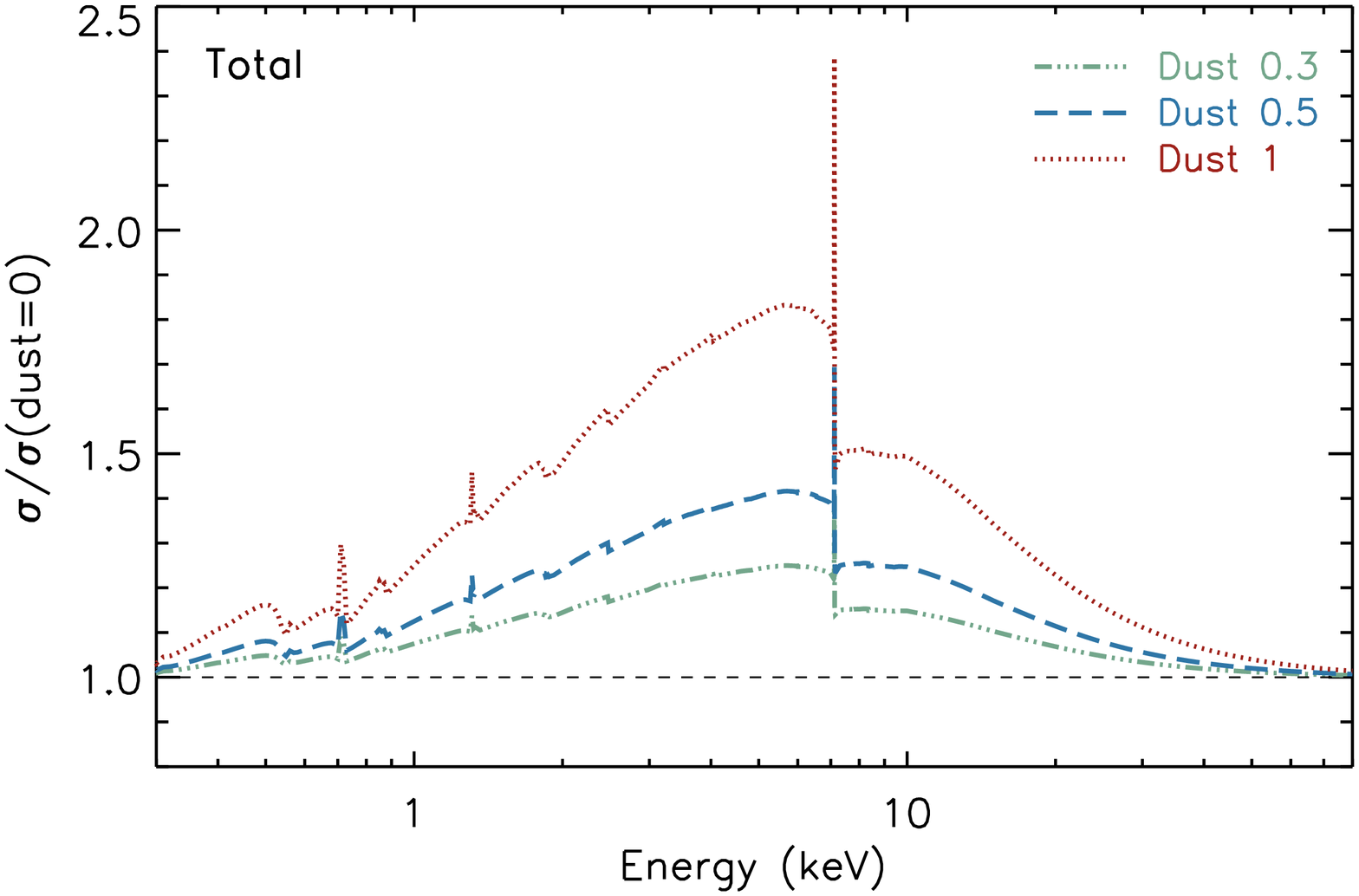} 
\includegraphics[width=0.49\textwidth]{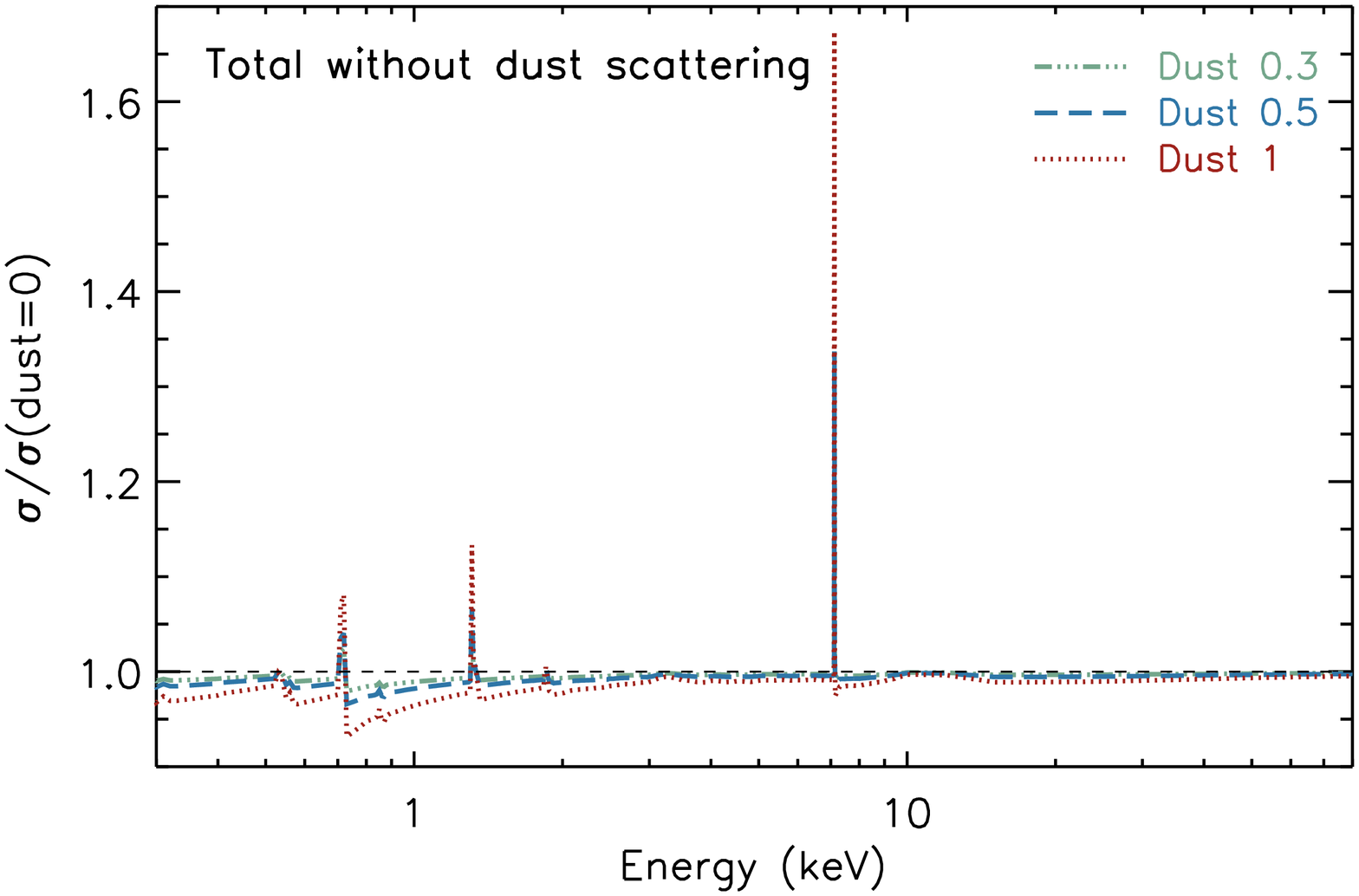} 
\caption{Ratio of the cross-sections with and without dust used in \reflex\ considering different dust depletions. {\it Top left panel:} ratio between the photoelectric cross section assuming different dust depletion and that of non-dusty gas.  {\it Top right panel:} ratio between the scattering cross-section (i.e., Rayleigh, Compton and dust) of dusty and non-dusty gas. {\it Bottom right panel:} ratio between the total cross-section of dusty gas and that of non-dusty gas. {\it Bottom left panel:} ratio between the total cross-section of dusty gas without dust scattering and that of non-dusty gas. The proto-solar abundances of \citet{Lodders:2009vs} were considered to calculate the cross-sections. The spectral resolution was set to 10\,eV.}
\label{fig:ratio_crosssections}
\end{figure*}

\begin{figure*}
\centering
\includegraphics[width=0.49\textwidth]{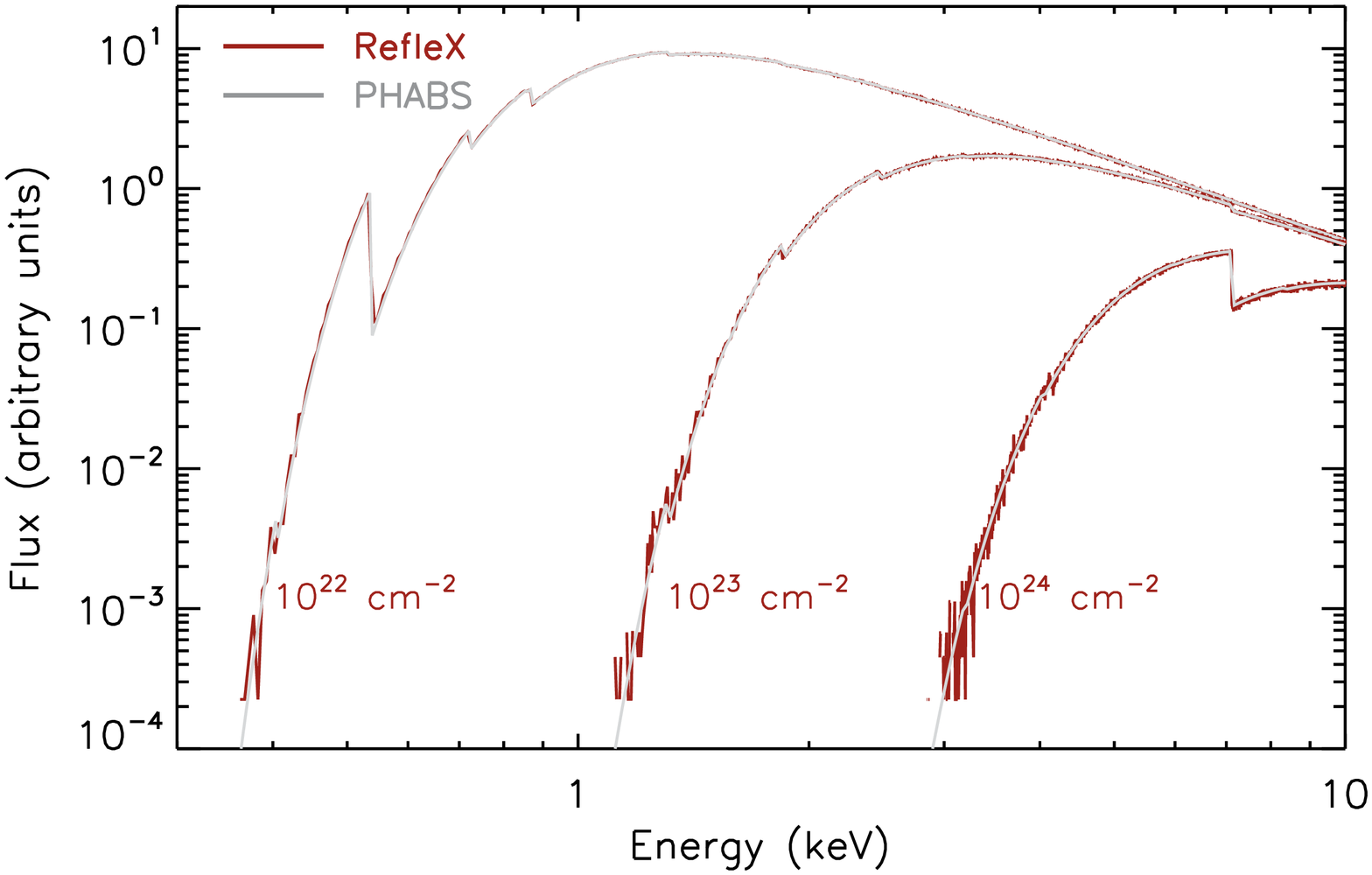} 
\includegraphics[width=0.49\textwidth]{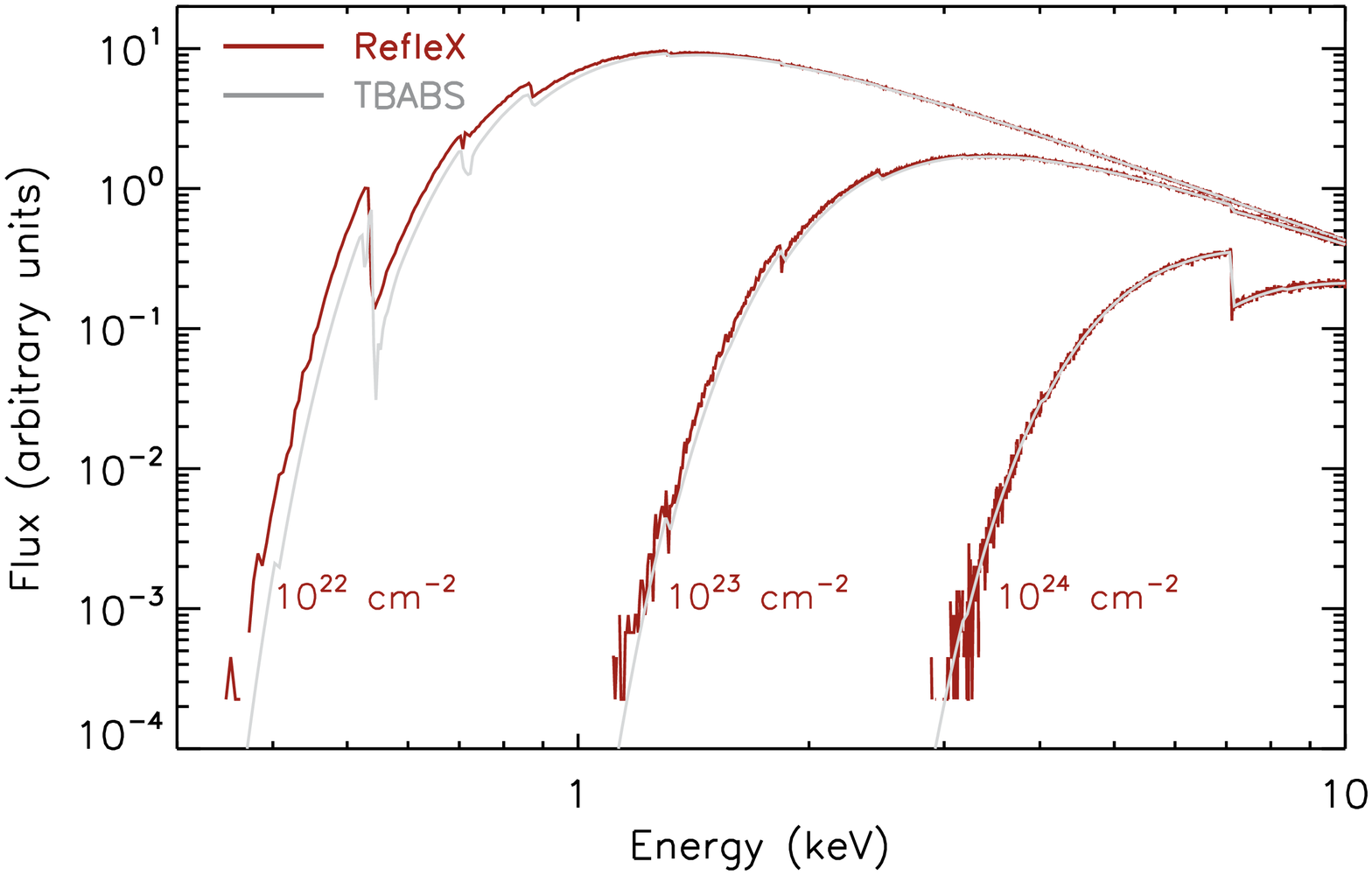} 
\caption{\textsc{RefleX} simulations versus the two most commonly used absorption models available in \textsc{xspec}. The simulations were carried out assuming that the absorbing material is distributed in a slab, considering different line-of-sight column densities ($10^{22}$, $10^{23}$ and $10^{24}\rm\,cm^{-2}$), and a spectral resolution of 5\,eV. {\it Left panel}: \textsc{RefleX} simulations (red lines) versus \textsc{phabs} (grey lines). For consistency, in our simulations dust was not considered, the molecular gas fraction was set to zero, and we turned off fluorescence, Rayleigh and Compton scattering. {\it Right panel}: \textsc{RefleX} simulations (red lines) versus \textsc{tbabs} (grey lines). For consistency, dust depletion was set to \textsc{dust 0.7}, while the molecular gas fraction to $f_{\rm H2}=20\%$, and the simulations ignored fluorescence, Compton scattering, as well as dust scattering.}
\label{fig:phabs_tbabs}
\end{figure*}
\subsection{Cross sections of dusty gas}\label{sec:crosssectionsdust}

We illustrate in Fig.\,\ref{fig:crosssections} the cross-sections of gas for different dust depletions. In the top left panel we show the photo-electric cross-section, which decreases as the dust depletion increases because of the increasing amount of metals in the dust grains. This effect is clearly illustrated in the top left panel of Fig.\,\ref{fig:ratio_crosssections}, which shows the ratio between the photoelectric cross-section of dusty and non-dusty gas, for different depletions. At the maximum dust depletion, the photoelectric cross-section decreases up to a factor $\sim 4$ with respect to the non-dusty gas case. The cross-section of dust is shown in the top-right panel of Fig.\,\ref{fig:crosssections} for three values of depletion, while in the bottom-left panel we show the dust cross-section without scattering. As expected, the cross-section increase with the dust depletion, as more metals are found in dust grains. Comparing the two figures it is clear that dust scattering contributes significantly to the overall dust cross-section.
In the bottom-right panel we show the total cross section of dusty gas for \textsc{dust 1} (black solid line), together with the contribution of different cross-sections (i.e., photoelectric, Rayleigh, Compton and dust), over a broad range of energies (0.3--200\,keV). Dust plays a strong effect below 10\,keV (see also top right panel of Fig.\,\ref{fig:ratio_crosssections}), and dominates the overall cross-section in the $\sim 3-10$\,keV range. Above $\sim$10\,keV Compton scattering dominates the interaction between photons and gas.

In the bottom right panel of Fig.\,\ref{fig:ratio_crosssections} we show the ratio between the total cross-section of dusty gas and that of non-dusty gas [$\sigma/\sigma(\textsc{dust 0})$]. The strong increase in the cross-section of dusty gas, with the ratio reaching $\sim 2.5$ around the Fe edge at $\sim 7$\,keV, is driven by the contribution of dust scattering. Since dust scattering is preferentially forward, and the scattering angles are typically rather small, this process does not lead to strong changes in the X-ray spectra, although some photons could be lost, while it could give rise to scattering halos (see \S\ref{sec:dusthalo}). The very large contribution of dust grains to the overall scattering cross section is shown in the top right panel of Fig.\,\ref{fig:ratio_crosssections}. To better assess the effect of dust we show, in the bottom left panel of Fig.\,\ref{fig:ratio_crosssections}, $\sigma/\sigma(\textsc{dust 0})$ when excluding dust scattering. The figure illustrates clearly the effect of shielding (\S\ref{sec:reflex_dust2}), which can lead to $\sigma/\sigma(\textsc{dust 0})<1$, particularly at energies $\lesssim 3$\,keV. This effect produces a decrease in the number of photons that are absorbed at low energy by dusty gas with respect to non-dusty gas. Another effect that is clear from the figure is the strong increase of the cross-section around the edges of elements that are found in dust (i.e., C, O, Si, Fe), due to the presence of NEXAFS (\S\ref{sec:reflex_dust2}).

\section{Comparison with photoelectric absorption models}\label{sec:reflexvstbabs}

\reflex\ was thoroughly tested in \citet{Paltani:2017fa}, where we compared the results of our simulations to several existing X-ray spectral models, considering different geometries of the reprocessing material, and discussed about the differences, which are due to the larger number of physical processes included into \reflex. In order to further validate this new version of the code, we briefly compare here the new version of \reflex\ to two of the most-used existing photoelectric absorption models within \textsc{XSPEC} \citep{Arnaud:1996kx}: \textsc{phabs} (\S\ref{sect:phabs}) ad \textsc{tbabs} (\S\ref{sect:tbabs}). In these simulations we assume a primary X-ray continuum in the form of a power-law with a photon index of $\Gamma=1.9$.

\subsection{PHABS}\label{sect:phabs}
We start by benchmarking our simulations with \textsc{phabs}, which considers photoelectric absorption from neutral gas. Therefore, in our \reflex\ simulations dust was not considered (\textsc{dust 0}), and the molecular gas fraction was set to zero (\textsc{h2fraction 0}). Moreover, to be consistent with the model, our simulations ignored fluorescence, Rayleigh and Compton scattering, which are otherwise included by default in \reflex. This was done by using a functionality of \reflex\ that allows to turn off some of the physical processes\footnote{\textsc{physics fluor off, physics rayleigh off, physics compton off}}. The abundance was set to be the same in both the \reflex\ simulations and the \textsc{xspec} model (\textsc{wilm}; \citealp{Wilms:2000vn}).
The results of our simulations (red line), together with those obtained using \textsc{phabs} (grey line), are shown in the left panel of Fig.\,\ref{fig:phabs_tbabs} for three different values of column density: $N_{\rm H}=10^{22}$, $10^{23}$ and $10^{24}\rm\,cm^{-2}$. The agreement between the \reflex\ simulations and the spectral model is perfect, regardless of the column density.

\subsection{TBABS}\label{sect:tbabs}

\textsc{tbabs} \citep{Wilms:2000vn} considers absorption by dust grains, and therefore dust depletion was set to \textsc{dust 0.7} in our \reflex\ simulations. To be consistent with the assumption of the model, we set the molecular gas fraction to $f_{\rm H2}=20\%$, and used the abundances from \citet{Wilms:2000vn}. Since \textsc{tbabs} does not consider fluorescence, Compton and dust scattering, we turned off all these physical processes\footnote{\textsc{physics fluor off, physics rayleigh off, physics compton off, physics dust\_scat off.}}. The comparison between our simulations and \textsc{tbabs} is shown in the right panel of Fig.\,\ref{fig:phabs_tbabs} for three different values of column density. While at $N_{\rm H}\sim 10^{23}-10^{24}\rm\,cm^{-2}$ there is an excellent agreement, our \reflex\ simulations show weaker absorption at $10^{22}\rm\,cm^{-2}$. This difference is associated to the different grain sizes considered in the two codes, which results in a different impact of shielding (\S\ref{sec:reflex_dust2}). The grain size adopted in \reflex\ follows \citeauthor{Draine:2003kb} (\citeyear{Draine:2003kb}; see also \citealp{Weingartner:2001eb}; \S\ref{sec:reflex_dust1}), while that of \textsc{tbabs} assumes grains with a size distribution following a power law with an index of $-3.5$ in the $0.025-0.25\mu$m size range \citep{Mathis:1977gx,Draine:1984ak}. Our choice to use the grain size distribution of \citet{Weingartner:2001eb} is due to the fact that their model is consistent with several different observables, such as i) the infrared emission from interstellar dust (e.g., \citealp{Li:2001ka}), ii) X-ray scattering halos  (e.g., \citealp{Rolf:1983rr}), and iii) optical and UV scattered light (e.g., \citealp{Draine:2003vd}).

\begin{figure}
\centering
\includegraphics[width=0.49\textwidth]{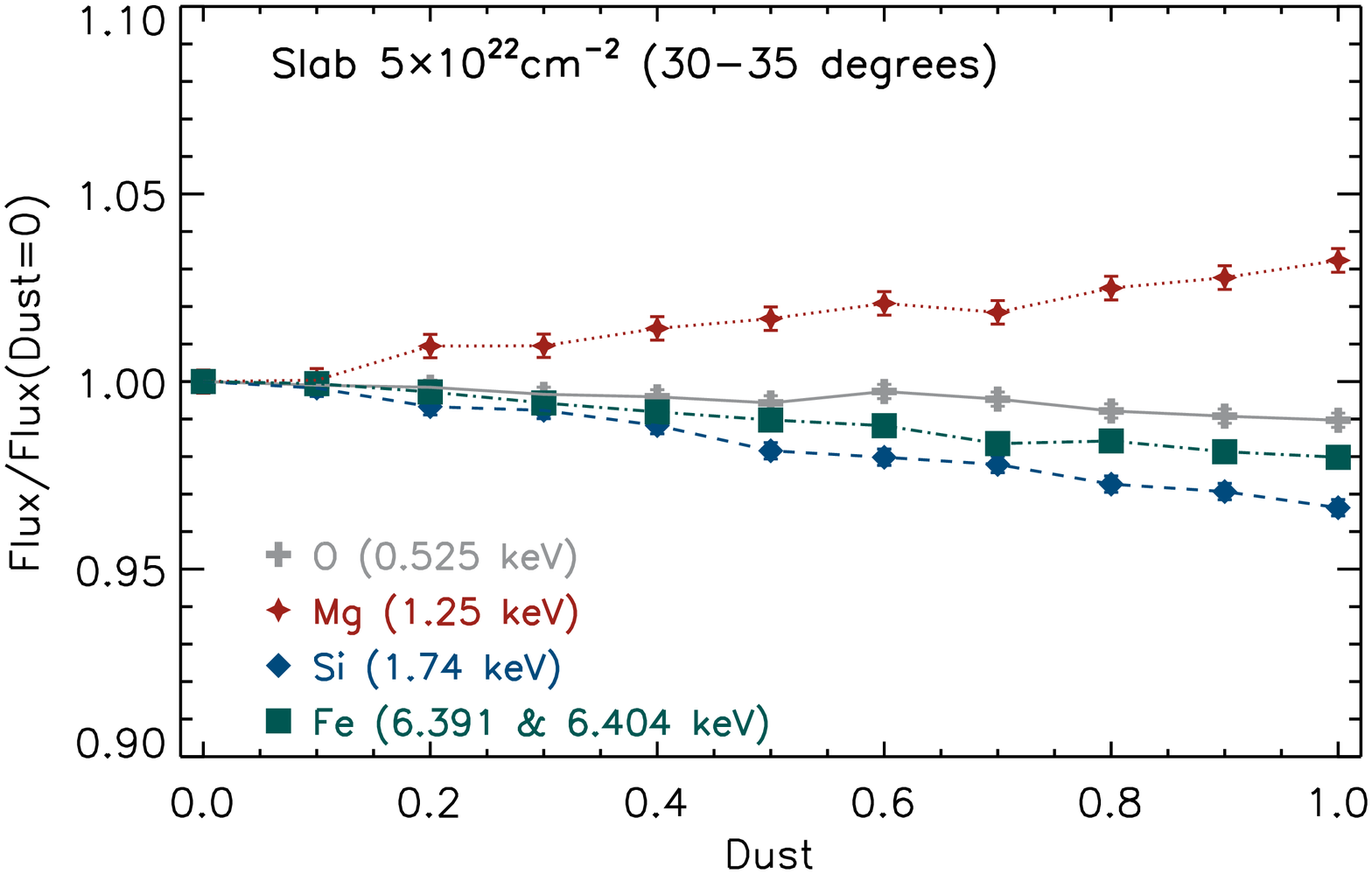} 
\includegraphics[width=0.49\textwidth]{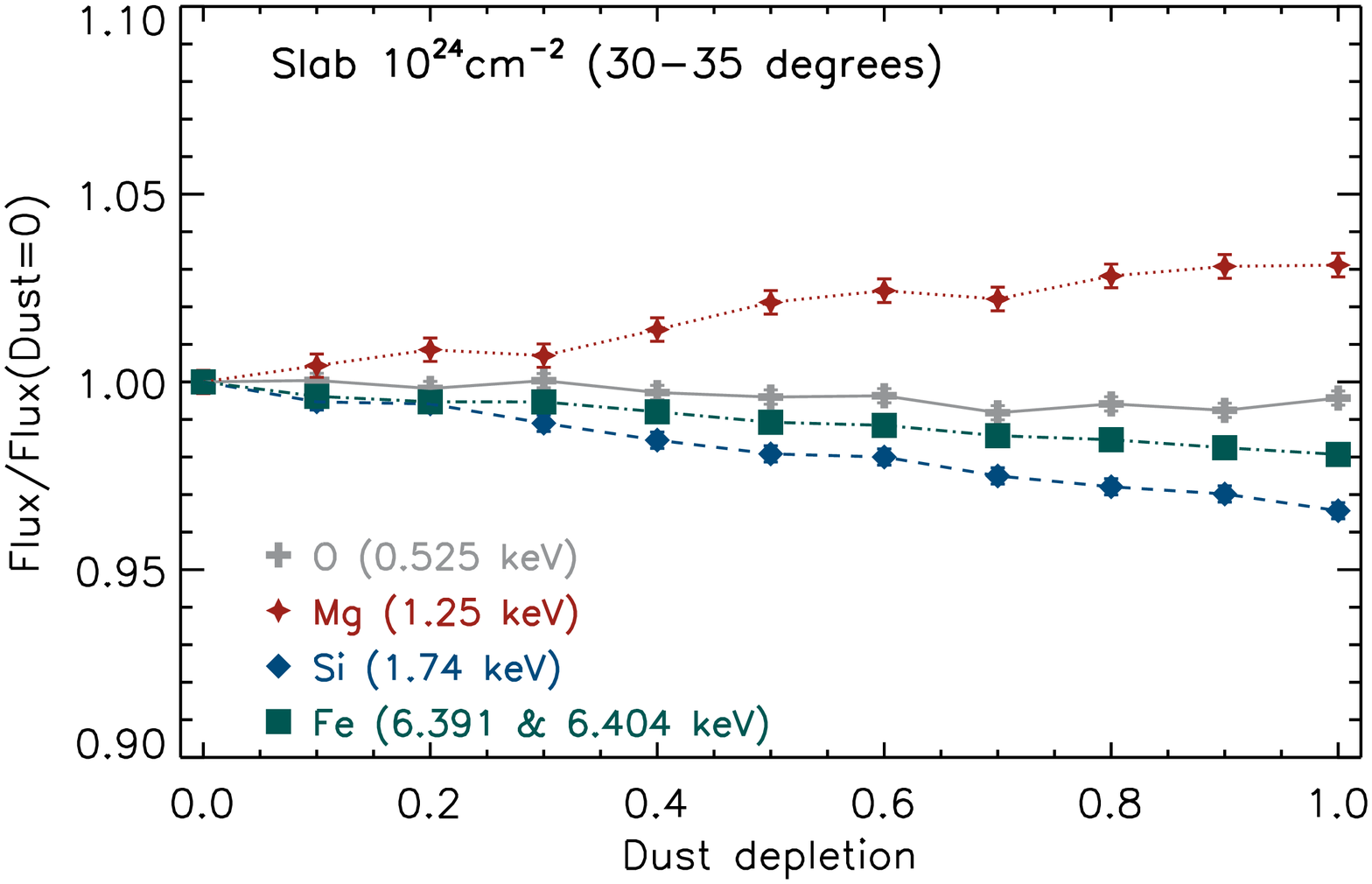} 
\includegraphics[width=0.49\textwidth]{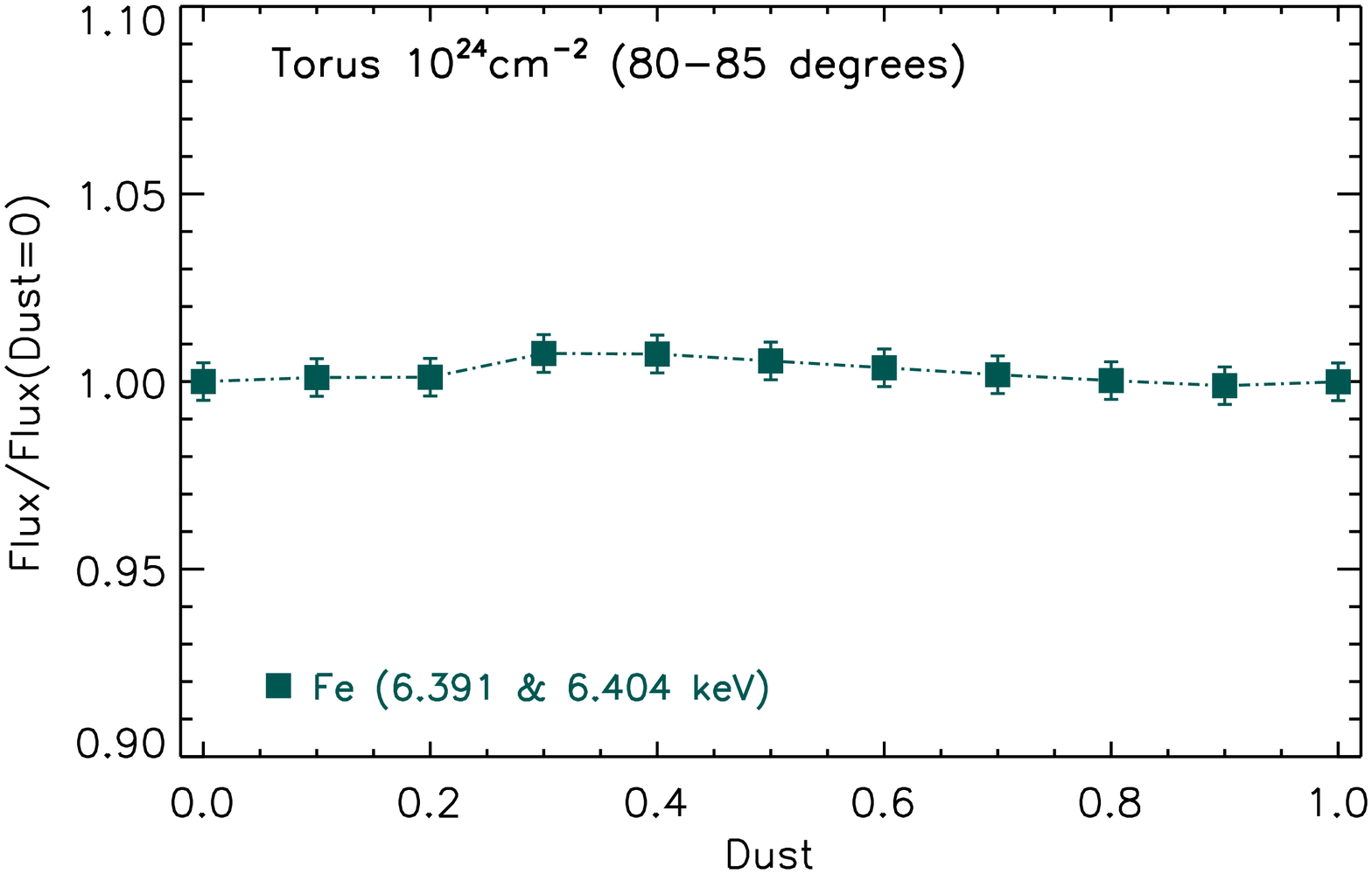} 
\caption{Ratio between the flux of fluorescence lines for \textsc{RefleX} simulations of dusty and non-dusty gas for different dust depletions, geometries and inclination angles. Fluorescence lines from some of the elements contained in the dust grains implemented in the latest version of \textsc{RefleX} are shown: oxygen ($0.525$\,keV), magnesium (1.25\,keV), silicon (1.74\,keV) and iron ($6.391$\,keV and $6.404$\,keV). Only photons that underwent at least a fluorescence were selected. {\it Top panel:} simulations for a slab with a vertical (i.e. along the shorter side) column density of $N_{\rm H}=5\times10^{22}\rm\,cm^{-2}$ observed at an inclination angle of $30^{\circ}$ with respect to the normal to the slab. {\it Middle panel:} same as the top panel for $N_{\rm H}=10^{24}\rm\,cm^{-2}$. {\it Bottom panel:} ratio between the flux of Fe produced in dusty and non-dusty gas for different values of dust depletion, assuming a torus with a column density $N_{\rm H}=10^{24}\rm\,cm^{-2}$ observed edge-on ($80^{\circ}$).}
\label{fig:fluorescentlines}
\end{figure}

\begin{figure}
\centering
\includegraphics[width=0.49\textwidth]{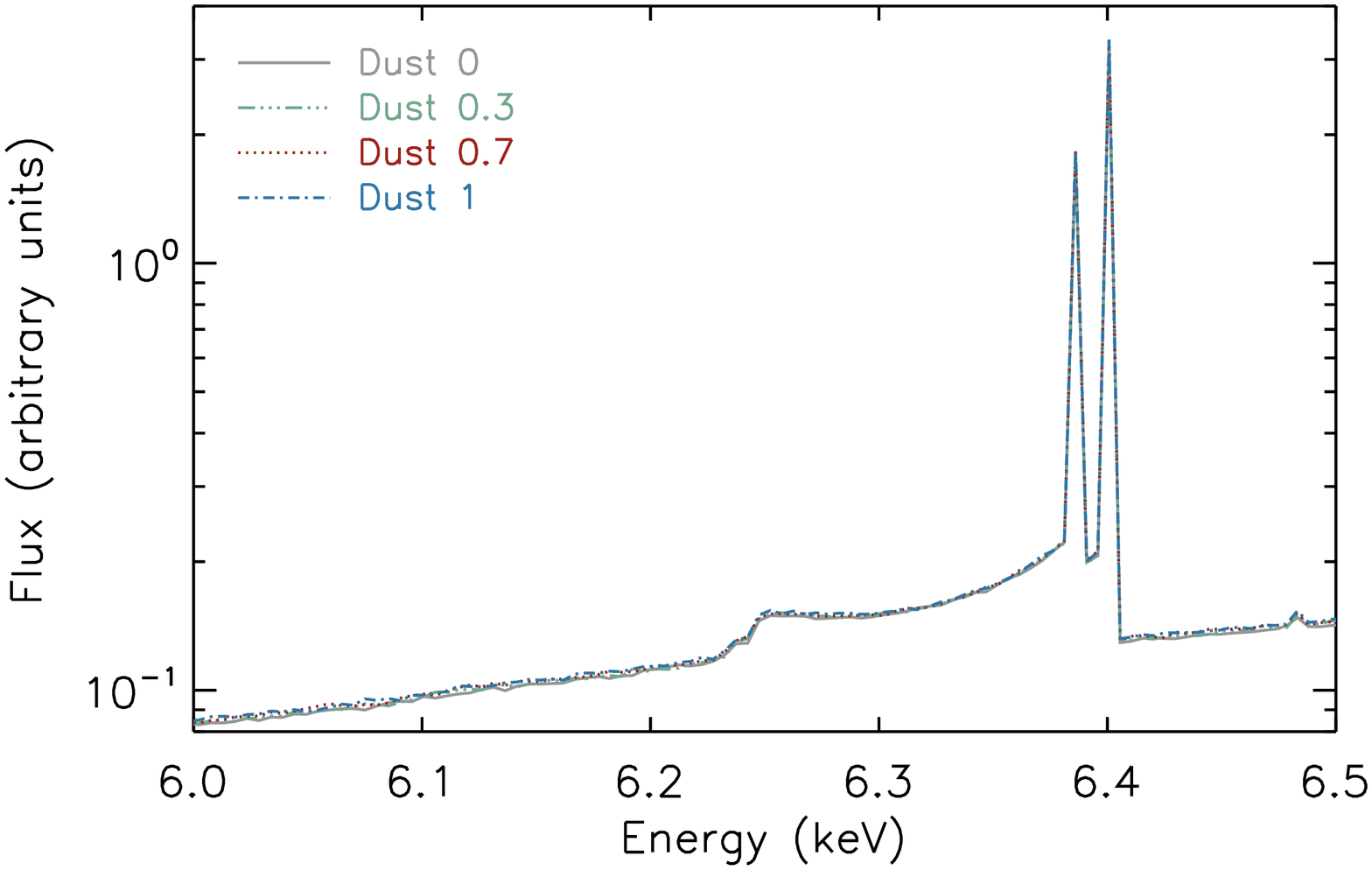} 
\includegraphics[width=0.49\textwidth]{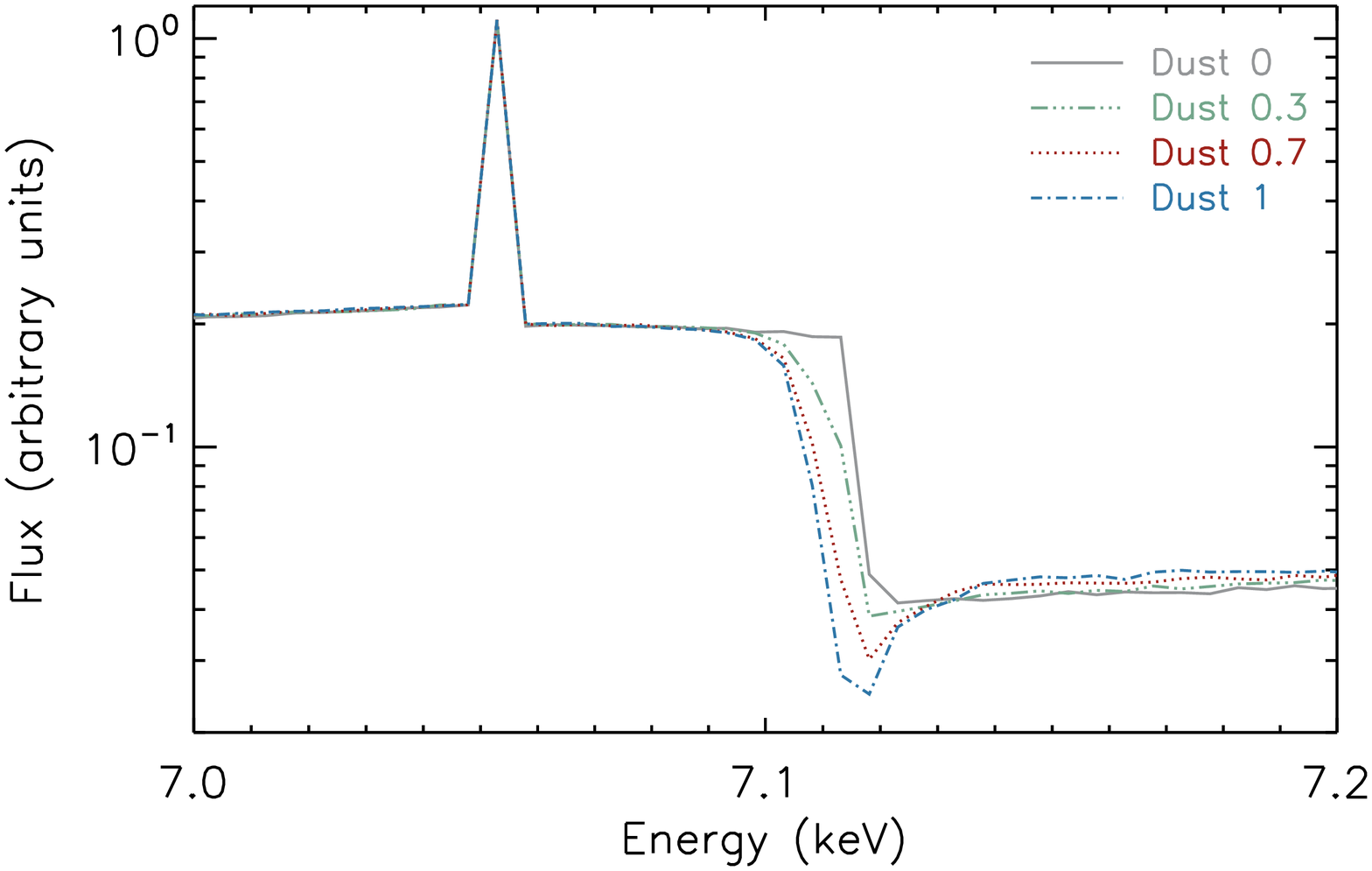} 
\caption{\textsc{RefleX} simulations around the Fe\,K$\alpha$ line and its Compton shoulder ($6.0-6.5$\,keV; top panel), and around the Fe\,K edge ($7.0-7.2$\,keV; bottom panel) for four different dust depletions assuming a torus geometry with a column density of $10^{24}\rm\,cm^{-2}$ observed edge-on ($85-90^{\circ}$), and considering a spectral resolution of 5\,eV.}
\label{fig:IronKregion}
\end{figure}

\section{X-ray emission and dusty gas}\label{sect:Xraydusty}

\subsection{The effect of dust on fluorescence lines}\label{sec:fluorescent}

We examine here the effect of dust on some fluorescence lines that are typically observed in astrophysical sources. We focus on lines that are produced by elements found in the dust grains we considered: oxygen ($0.525$\,keV), magnesium (1.25\,keV), silicon (1.74\,keV) and iron ($6.391$\,keV and $6.404$\,keV). We started by assuming a simple scenario with a point-like X-ray source located at $10^{17}\rm cm$ from a slab with a length of $2\times 10^{25}$\rm cm and a width of $10^3$\,cm. This configuration is used to reproduce a situation in which the X-ray source is located close to an extended cloud of dusty gas, and is similar to that assumed in the \textsc{pexrav} model \citep{Magdziarz:1995ic} that has been widely used for active galactic nuclei (AGN). We considered two different densities of the slab ($5\times10^{19}\rm\,cm^{-3}$ and $10^{21}\rm\,cm^{-3}$), which resulted into vertical column densities of the disk (i.e., along its shorter side) of $5\times 10^{22}\rm\,cm^{-2}$ and $10^{24}\rm\,cm^{-2}$, respectively. The X-ray continuum was generated to have a power-law spectrum with $\Gamma=1.8$ and a cutoff energy of 200\,keV (\textsc{cutoff} in \reflex ), consistent with the typical spectrum of nearby AGN (e.g., \citealp{Winter:2009iq,Ricci:2017if,Ricci:2018mp,Balokovic:2020fc}). We considered only photons that underwent at least one fluorescence event in the simulations, selected those observed with an inclination angle of $30-35^{\circ}$ with respect to the normal to the slab, and integrated the flux around the emission lines. In the case of the Fe\,K$\alpha$ photons we integrated in the range between 6.38 and 6.41\,keV to exclude photons that are down-scattered from the Fe\,K$\alpha$ doublet into the Compton shoulder. In all simulations we used the abundance table of \citeauthor{Wilms:2000vn} (\citeyear{Wilms:2000vn}; \textsc{wilm} in \reflex). The results are shown in the top and middle panels of Fig.\,\ref{fig:fluorescentlines}, and highlight the small effect of dust on the flux of the fluorescence lines, with the difference with respect to the dust-free simulations being at most $\sim 5\%$. In particular, the difference is $\lesssim 1\%$  for iron and oxygen.

We also tested a different scenario, in which the X-ray source is surrounded by a torus with an equatorial column density of $N_{\rm H}=10^{24}\rm\,cm^{-2}$, a covering factor of 0.7 and observed edge-on. This was done collecting photons in the $80-85^{\circ}$ range, and applying the same selection criteria we used for the slab, collecting both absorbed and reflected X-ray photons. The results are shown for the Fe\,K$\alpha$ line in the bottom panel of Fig.\,\ref{fig:fluorescentlines}.
In Fig.\,\ref{fig:IronKregion} we illustrate the X-ray spectra around the Fe\,K$\alpha$ line (top panel) and Fe\,K edge (bottom panel) for different values of the \textsc{dust} parameter. No clear difference is observed in the Fe\,K$\alpha$ doublet, nor in its Compton shoulder. On the other hand, a clear difference between the X-ray emission produced in dusty and dust-free gas is found around the Fe\,K edge, due to the strong effect of the NEXAFS (see \S\ref{sec:reflex_dust2}). As expected, the difference increases with the dust depletion, since more iron atoms are found in dust.

\begin{figure}
\centering
\includegraphics[width=0.43\textwidth]{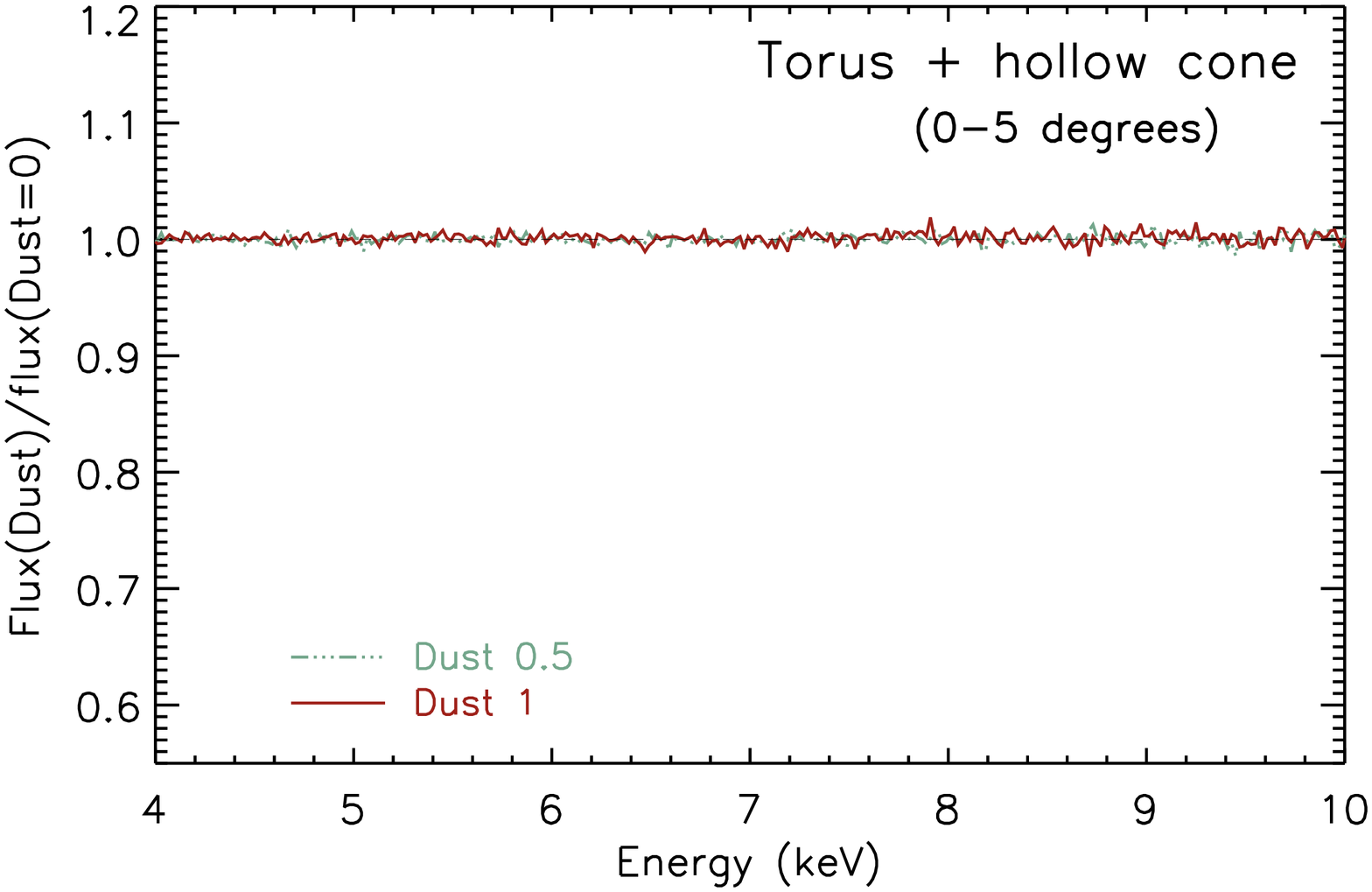} 
\includegraphics[width=0.43\textwidth]{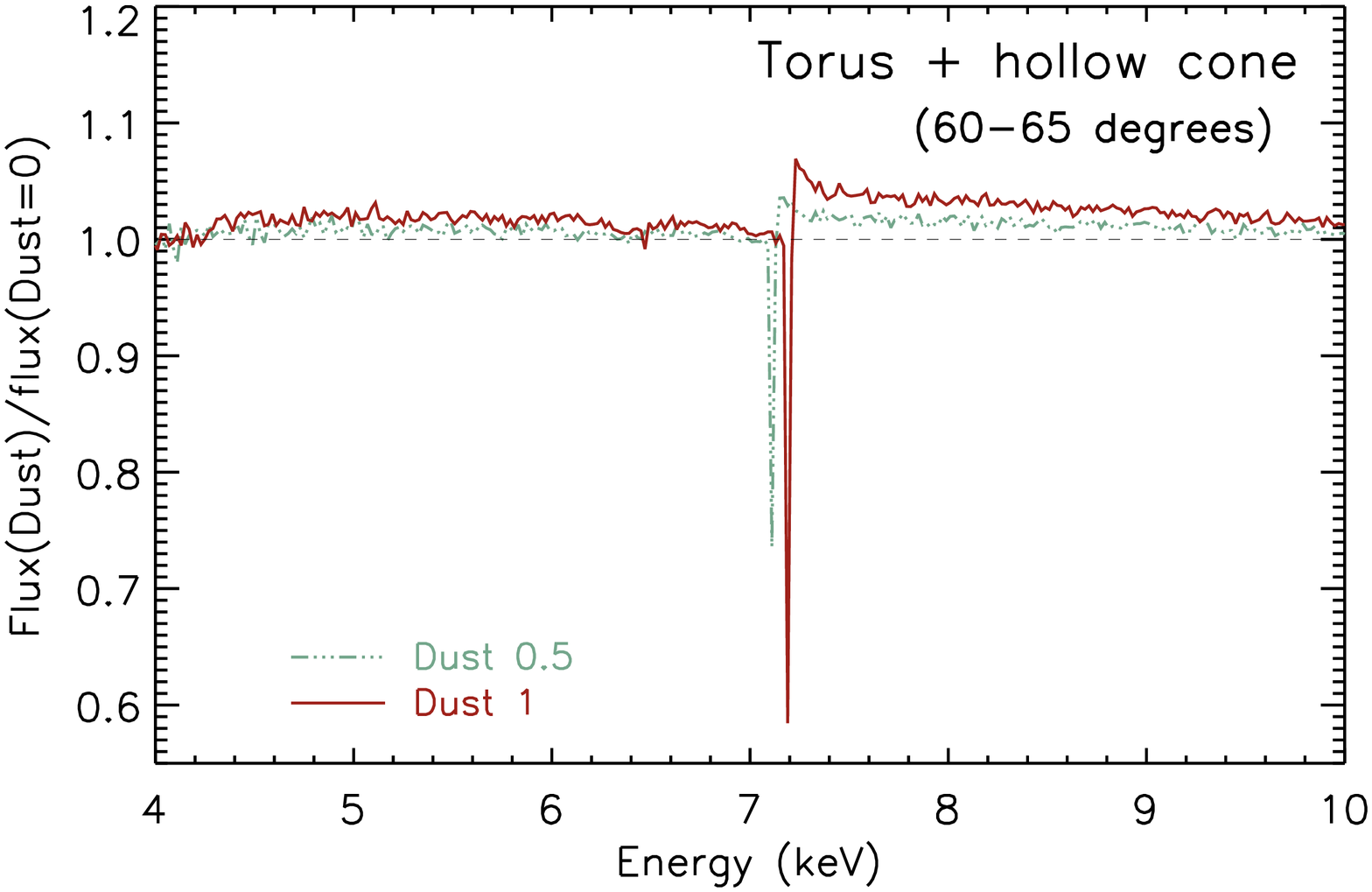} 
\includegraphics[width=0.43\textwidth]{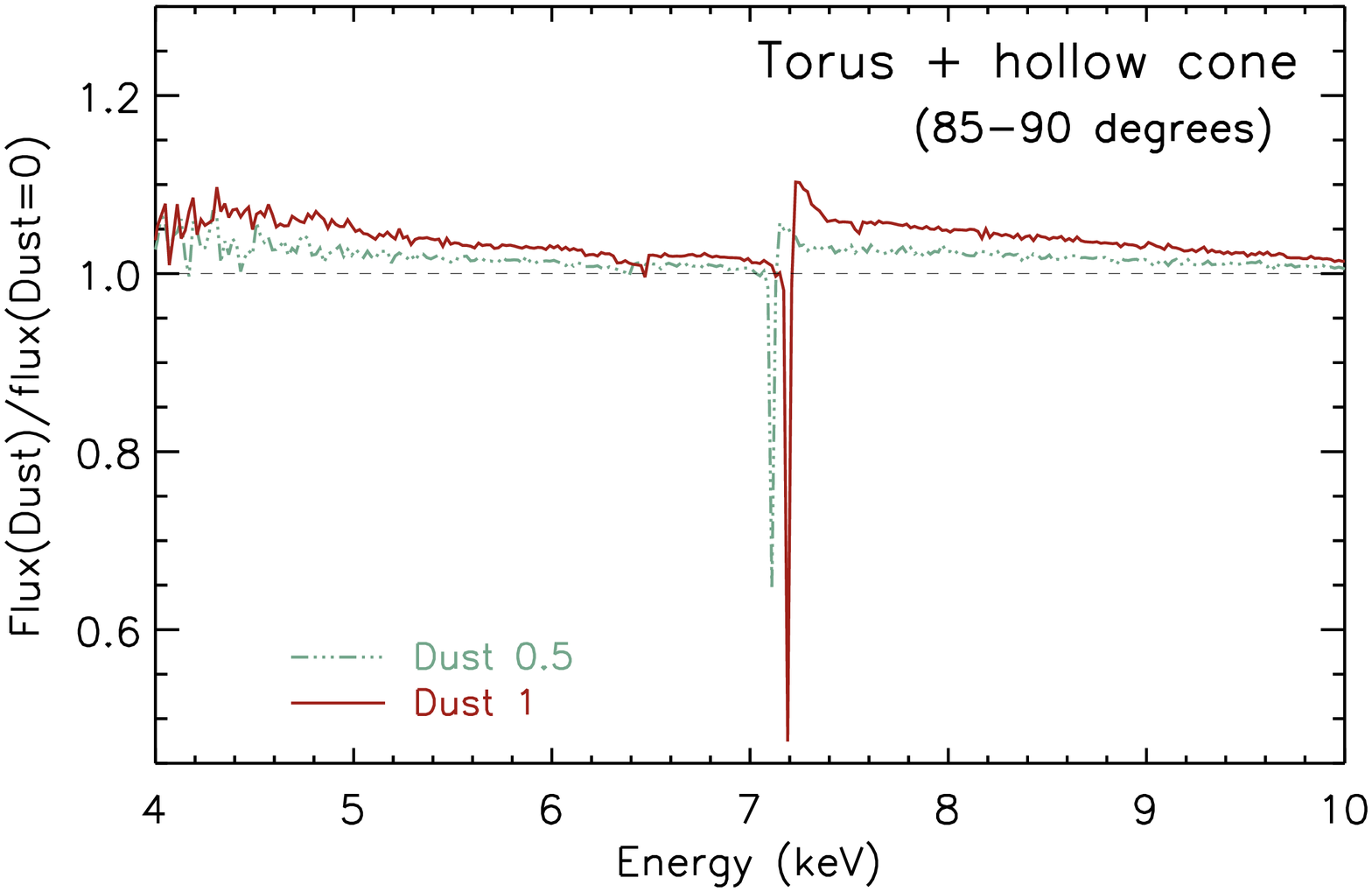} 
\includegraphics[width=0.43\textwidth]{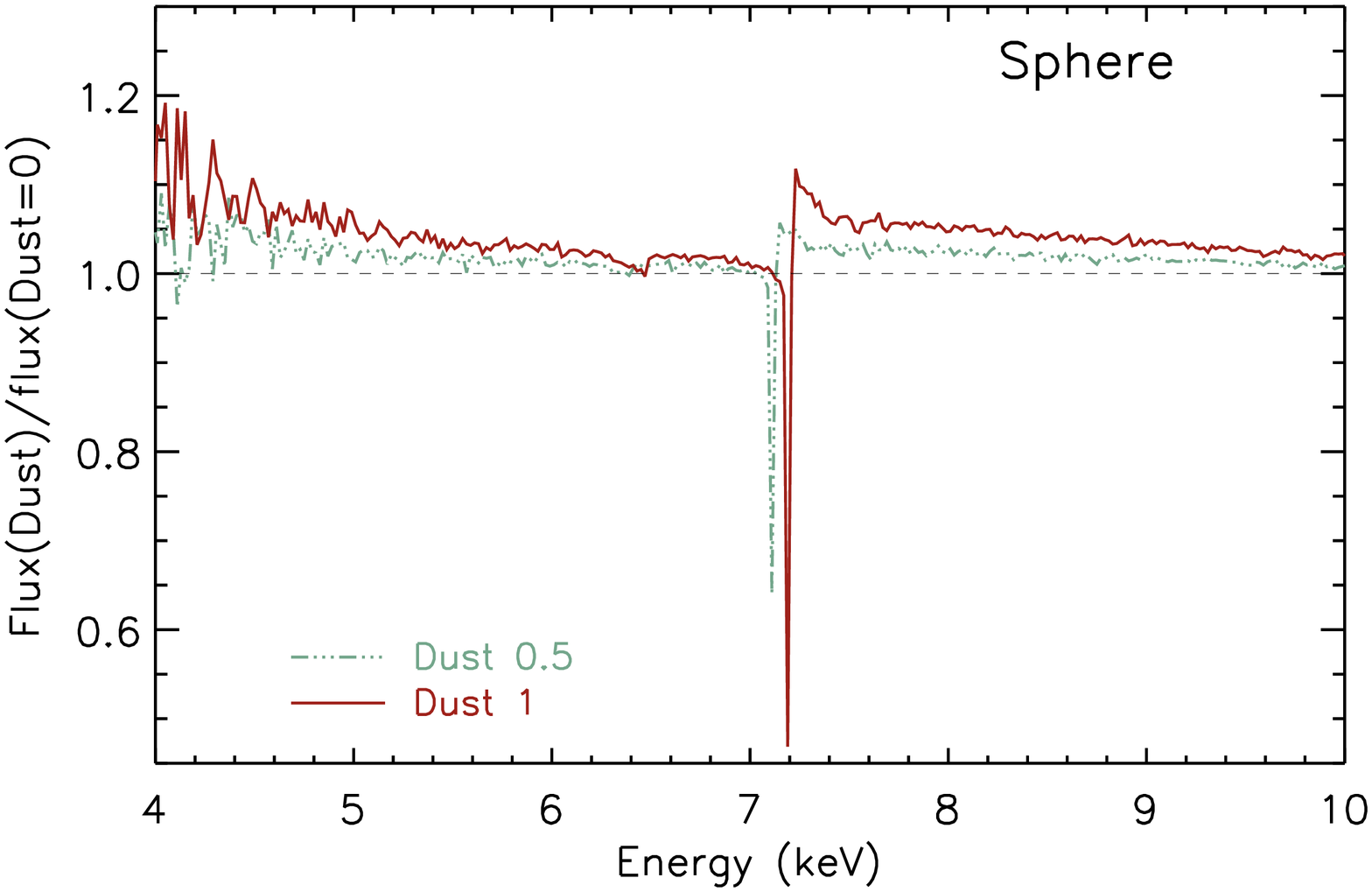} 
\caption{Ratio between our \textsc{RefleX} simulations with and without dust, for different dust depletions: 0.5 (green dot-dot-dashed lines) and 0.7 (red continuous lines). The spectral resolution was set to 20\,eV. The geometrical setup of these simulations includes a torus plus a hollow cone (top three panels) and a sphere (bottom panel). For the torus plus hollow cone geometry the spectra were obtained considering different inclinations: pole-one ($0-5^{\circ}$), intermediate ($60-65^{\circ}$) and edge-on ($85-90^{\circ}$). For the spherical geometry the photons were also selected considering an angle of $5^{\circ}$. The simulations with \textsc{dust 1} were shifted by 80\,eV for visual clarity.}
\label{fig:ratiosmodels}
\end{figure}

\clearpage

\subsection{Comparison with spectroscopical simulations ignoring dust}\label{sect:simulations}

Here we examine the effect on the broad-band X-ray spectra of astrophysical sources of considering or not dusty gas, by simulating complex geometries with \textsc{RefleX}. We explored a few different geometries that are very relevant to AGN, but our results can be extended to any type of source of X-ray radiation. In the simulations we considered the same parameters for the X-ray continuum (i.e., photon index and high-energy cutoff) reported in \S\ref{sec:fluorescent}, and a molecular fraction of $f_{\rm H2}=20\%$. We tested also larger values of the molecular fraction ($f_{\rm H2}=80\%$; \citealp{Willingale:2013fm}) and found that this has only a marginal effect on our simulations. We started by considering a geometry that includes a torus and a hollow cone, i.e. the dusty components used by \cite{Andonie:2022rs} for the ad-hoc X-ray spectral model built with \reflex\ for the AGN in the Circinus Galaxy, based on high-resolution mid-IR observations \citep{Stalevski:2017fr,Stalevski:2019ej}. The column density of the torus was set to $1.5\times10^{24}\rm\,cm^{-2}$, while the covering factor to 70\%, consistent with what is typically found in local AGN (e.g., \citealp{Burlon:2011yn,Ricci:2015fk,Ricci:2017if}). The hollow cone extends from 0.1 to 40\,pc, and has a maximum column density of $3\times 10^{22}\rm\,cm^{-2}$. The ratios between simulations performed assuming this geometry and considering dusty and non-dusty gas are shown in Fig.\,\ref{fig:ratiosmodels} for different inclination angles and dust depletions. For unobscured lines-of-sight (top panel), the simulations produce consistent results for dusty and non-dusty gas. A clear difference emerges for obscured lines-of-sight (middle panels), particularly around the Fe\,K edge, and it increases with the line-of-sight $N_{\rm H}$. For the most obscured and inclined scenarios the ratio between the X-ray spectra of dusty and non-dusty gas shows a dip at $\sim 60\%$ and an excess of up to $\sim 10\%$.

We also explored a different scenario, in which an X-ray source is surrounded by a dense sphere of dusty gas. Such a situation can be found at the center of Luminous and Ultra-luminous IR galaxies (U/LIRGs), where accreting SMBHs are typically completely embedded in a large amount of dusty gas (e.g., \citealp{Koss:2016zl,Ricci:2017aa,Ricci:2021wb,Yamada:2021vk}), and possibly in Compact Nuclei (e.g., \citealp{Aalto:2015ni,Aalto:2019yr,Falstad:2021wf}). We considered that the material is distributed homogeneously in a sphere with a radius of $10^{17}\rm\,cm$ and a density of $1.5\times 10^{7}\rm\,cm^{-3}$, which results in a column density of $1.5\times 10^{24}\rm\,cm^{-2}$. We selected photons in an interval of $5^{\circ}$ and the results are illustrated in the bottom panel of Fig.\,\ref{fig:ratiosmodels}. The ratio shows the same features observed for the edge-on case of the torus plus hollow cone geometry.

\subsection{Dust scattering halos}\label{sec:dusthalo}
Dust grains strongly increase the scattering cross section of gas (see the top right panel of Fig.\,\ref{fig:ratio_crosssections}). The small-angle scatter of X-ray photons off dust in the ISM can produce diffuse emission around a X-ray source (e.g., \citealp{Overbeck:1965ow,Hayakawa:1970oy,Martin:1970nh,Martin:1970hk}). The first of such halos was detected by \cite{Rolf:1983rr} studying {\it Einstein Observatory} observations of GX339$-$4. Further observations, carried out with the {\it Einstein Observatory} detected scattering halos around several additional X-ray sources (e.g., \citealp{Catura:1983zu,Mauche:1986vs,Mauche:1989xp}). Following observations with {\it ROSAT} (e.g., \citealp{Predehl:1995ga}), {\it XMM-Newton} (e.g., \citealp{Costantini:2005bx,Tiengo:2010rx,Pintore:2017sk,Jin:2017mq}) and {\it Chandra} (e.g., \citealp{Tan:2004bs,Corrales:2015ew,Corrales:2017hm}) detected several more dust scattering halos around Galactic sources. Dust scattering was also recently observed around the brightest gamma-ray burst detected so far (GRB 221009A; \citealp{Tiengo:2022ul}).
In some variable X-ray sources, the interaction between X-ray photons produced in repeated bursts of radiation and the dust grains can give rise to scattering echoes, that can be observed as rings (e.g., \citealp{Corrales:2019pr,Lamer:2021no}). Such structures have been found around variable Galactic sources (e.g., \citealp{Heinz:2015ms}). A rapid and strong decrease of the X-ray flux in AGN could give rise to ``ghost halos" \citep{Corrales:2015jy}, with the dust scattering halo becoming detectable in the X-rays. These halos typically contain few tens of percent of the X-ray flux at a few keV \citep{Predehl:1995ga}, and could be used to infer some fundamental properties of dust in the ISM (e.g., \citealp{Mauche:1986vs,Mathis:1991wa,Corrales:2015jy}).

\begin{figure}
\centering
\includegraphics[width=0.4\textwidth]{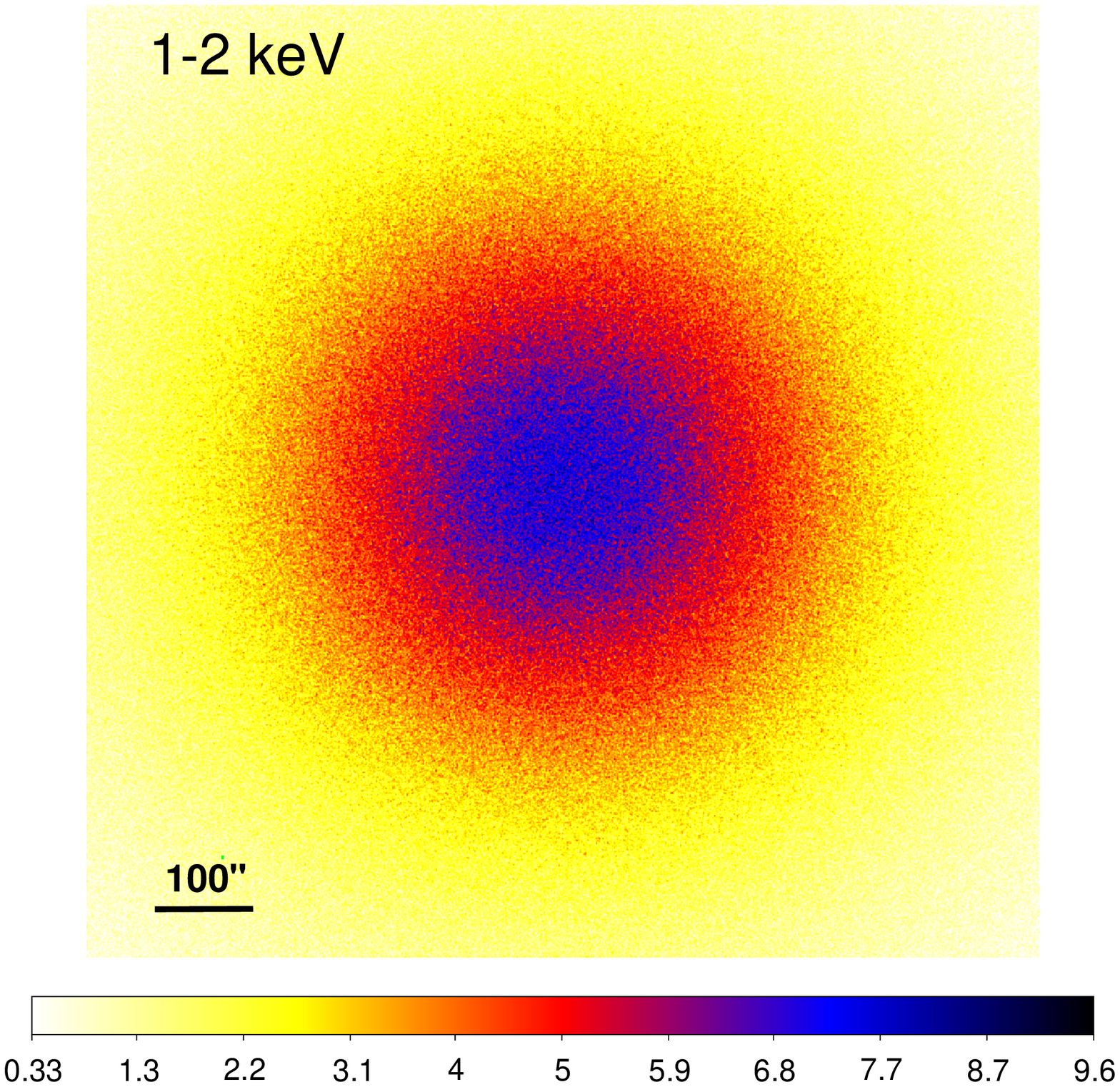}
\includegraphics[width=0.4\textwidth]{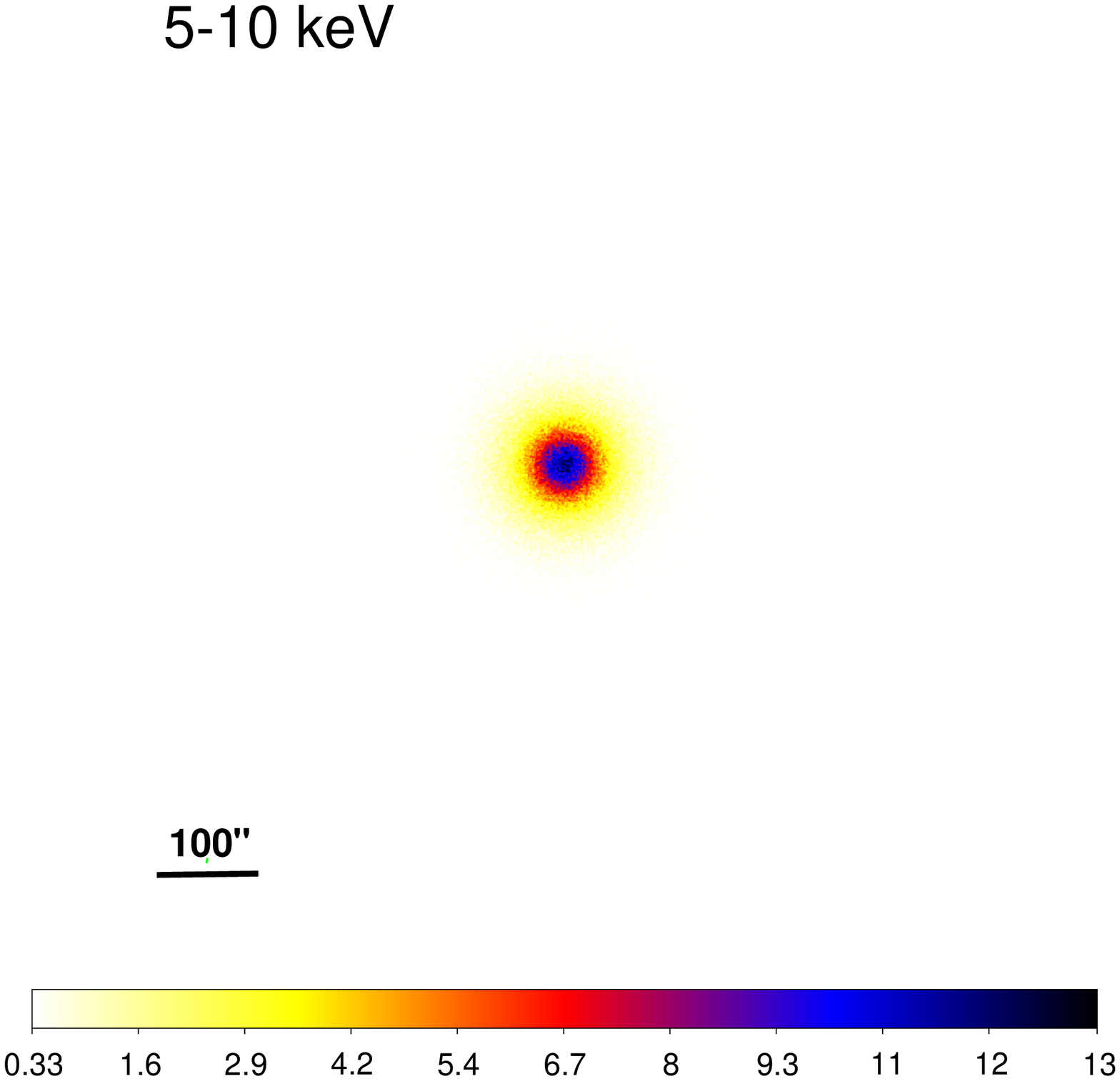}
\par\smallskip
\includegraphics[width=0.48\textwidth]{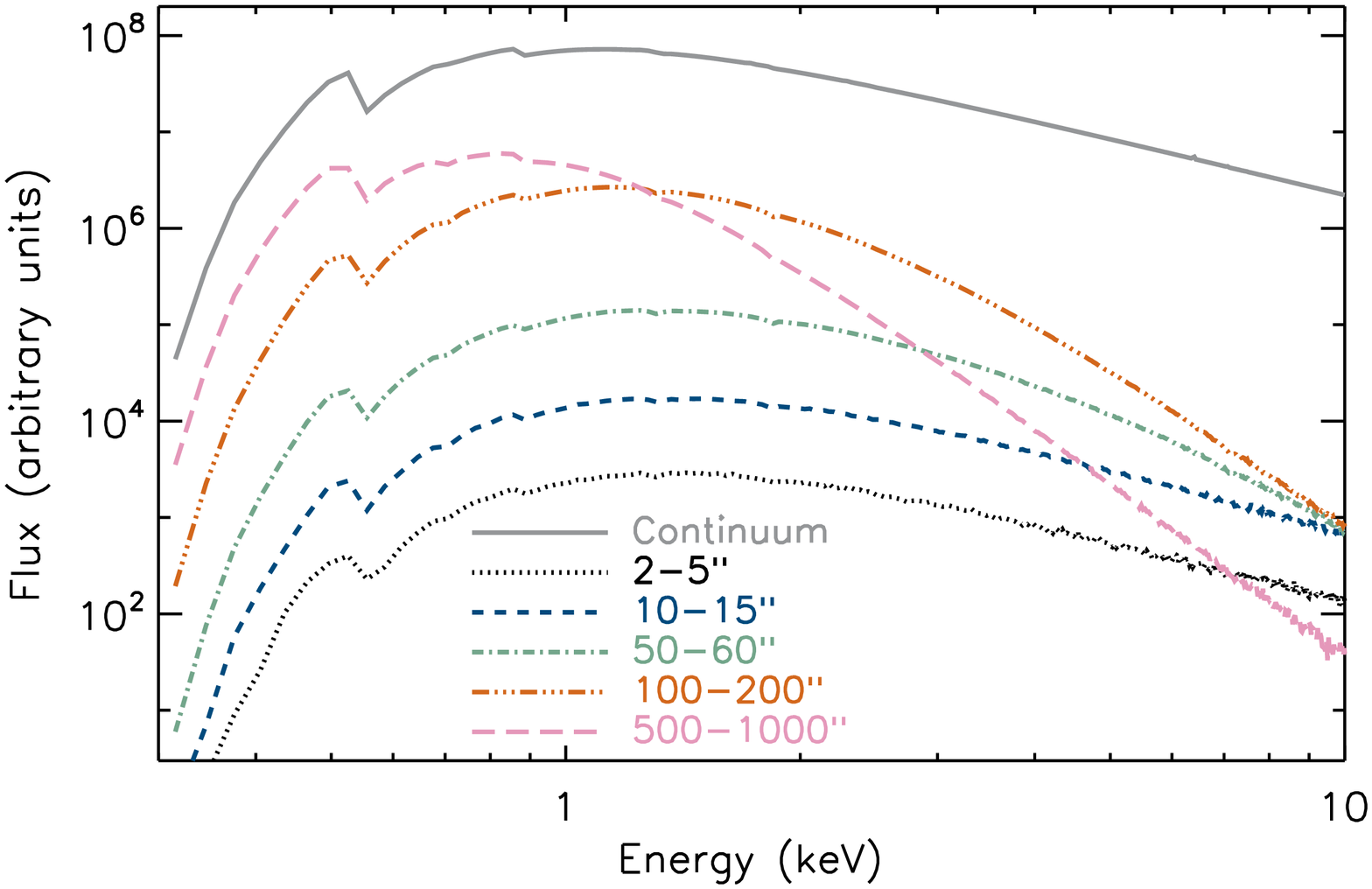}
\caption{Simulations of a dust scattering halo around an X-ray source. The simulations consider a mono-directional X-ray source and slab of dusty gas ($N_{\rm H}=10^{22}\rm\,cm^{-2}$; \textsc{dust 0.7}), and have a resolution of $0.54\arcsec$. The top and middle panel show images obtained in the 1--2\,keV and 5--10\,keV range, respectively. Only scattered photons were selected. The bottom panel shows the spectra extracted using different apertures centered on the source. The spectral simulation was set to 30\,eV.}
\label{fig:scatteringhalo}
\end{figure}

The imaging capabilities of \reflex\ allow the user to create images of dust scattering halos around astrophysical sources, and the comparison of \reflex\ simulated X-ray images with observations could be used to improve our understanding of the ISM responsible for the scattering halos (e.g., \citealp{Draine:2006qa,Xiang:2011qm}). While currently \reflex\ only considers one dust size distribution, it allows the user to set complex geometrical configurations of the dusty gas. As an example of the imaging capabilities of \reflex\, in the top and middle panel of Fig.\,\ref{fig:scatteringhalo} we show two images produced assuming a point-like X-ray source absorbed by a slab of dusty gas with a column density of $N_{\rm H}=10^{22}\rm\,cm^{-2}$, considering the abundances of \citet{Lodders:2009vs}, \textsc{dust 1} and a molecular fraction of $f_{\rm H2}=20\%$. The scattered photons were selected in the 1--2\,keV and 5--10\,keV range for the top and bottom panel, respectively. The figures show the scattering halos, that can extend to several hundreds arcseconds from the X-ray source, and the remarkable difference in the size of the scattering halos, which decreases with increasing energy. This is consistent with what expected, according to \cite{Draine:2003kb}, for $E\gtrsim 0.5$\,keV the median scattering angle can be approximated, for the \cite{Weingartner:2001eb} dust, by:
\begin{equation}\label{eq:scatteringangle}
\theta_{\rm s, 50}\sim 360\arcsec \left(\frac{\rm keV}{E}\right).
\end{equation}
Simulated images using the same configuration, but assuming a dust-free ISM, would show almost no scattered photon around the X-ray source. This is due to the smaller values of the Compton and Rayleigh cross-sections (see top right panel of Fig.\,\ref{fig:ratio_crosssections}), and to the fact that the angular distributions of Compton and Rayleigh scattering are more isotropic than that of Mie scattering. In the bottom panel of Fig.\,\ref{fig:scatteringhalo} we show the spectra obtained by extracting the flux using annuli centred on the X-ray source with different apertures. For the annuli with the smallest radii ($2-5''$ and $10-15''$, black dotted and blue dashed lines, respectively) the spectral shape is very similar to that of the X-ray source (grey continuous line). At larger radii instead the spectral shape starts to deviate from that of the X-ray continuum, becoming increasingly softer as the distance increases (green dot-dashed, red dot-dot-dashed and purple dashed lines for $50-60''$, $100-200''$ and $500-1000''$, respectively). The change in the spectral shape is due to the relation between the median scattering angle and the energy (Eq.\,\ref{eq:scatteringangle}), with softer photons being more easily scattered at larger angles. This softening of the dust-scattered radiation has been observed in X-ray binaries: for the giant dust ring found at $\sim 1^{\circ}$ from MAXI\,J1348$-$630, \cite{Lamer:2021no} inferred a much softer spectrum ($\Gamma \simeq 4$) with respect to that expected for the source during the high/soft state ($\Gamma\simeq 2$).

Thanks to its new functionalities, \textsc{RefleX} could be used to produce X-ray spectral models for a given aperture to fully reproduce the emission of astrophysical sources  (see \citealp{Corrales:2016bq,Smith:2016ad}). Moreover, \reflex\ could also be used to reproduce dust rings around variable X-ray sources, and to infer the spectral properties of the emission during the outbursts that give rise to the dust rings.

\subsection{RXTorusD: a dusty torus model}\label{sec:RXTorusD}

\subsubsection{The model}
\begin{figure*}
\centering
\includegraphics[width=0.46\textwidth]{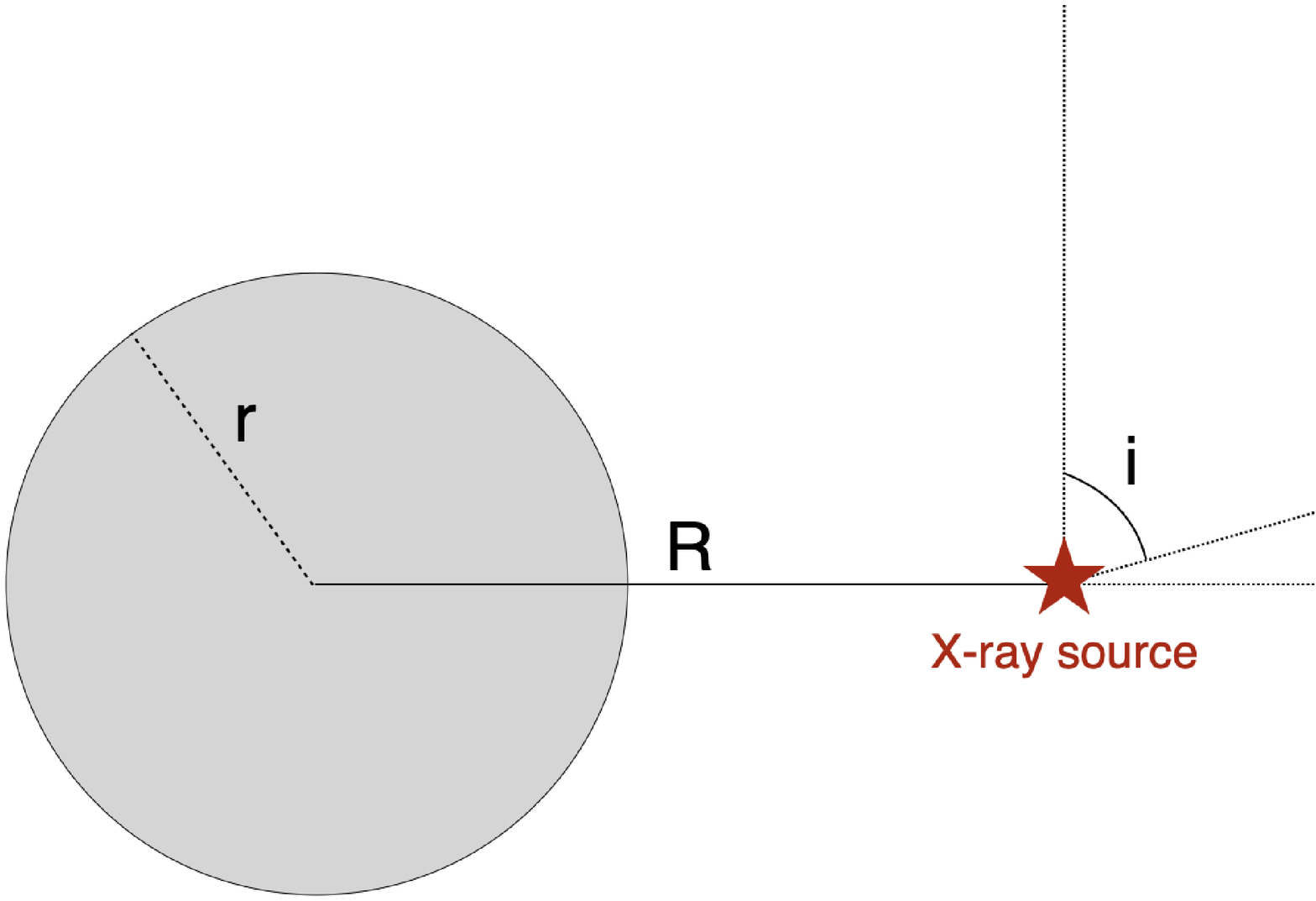}
\includegraphics[width=0.46\textwidth]{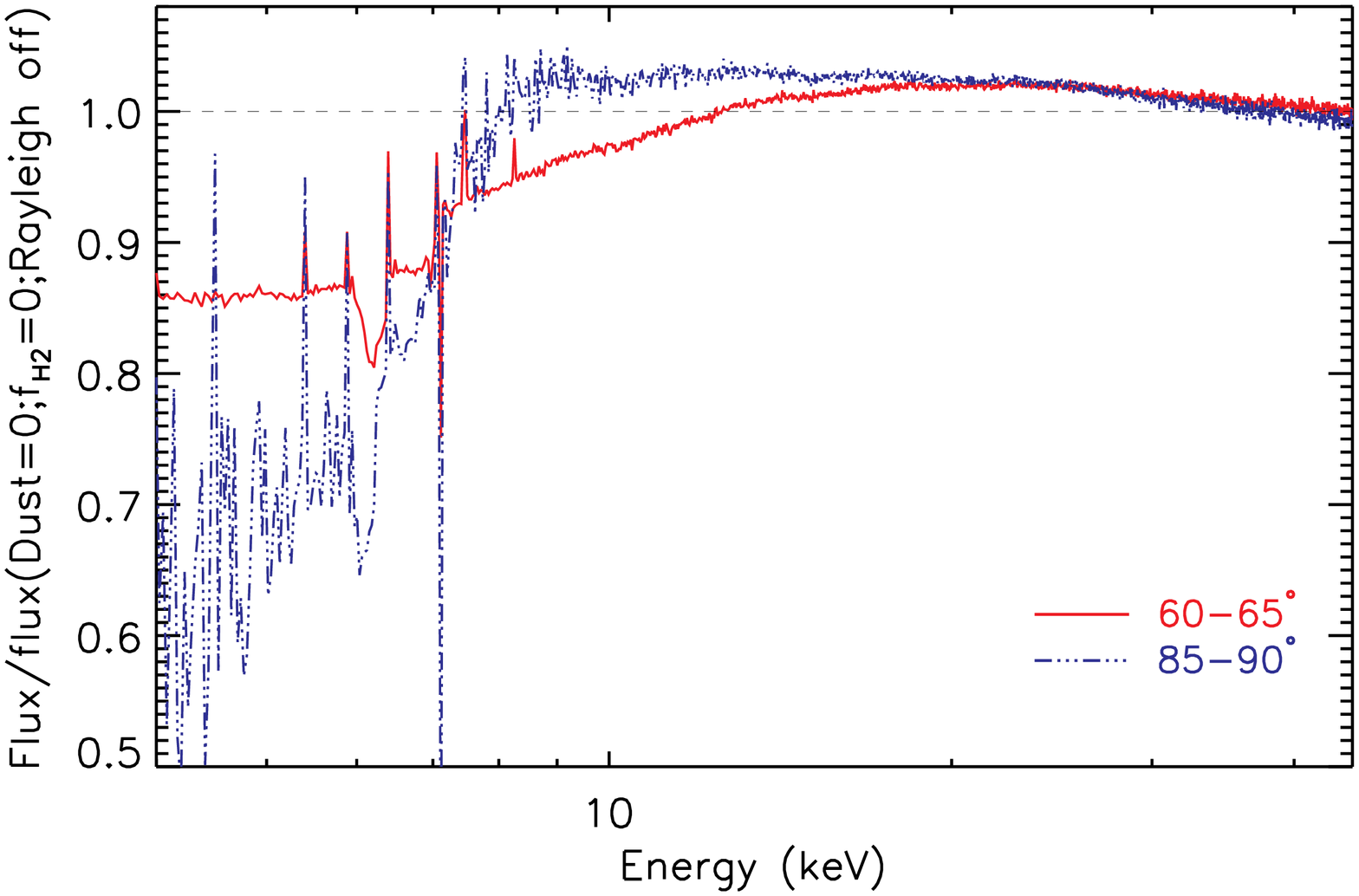}
\caption{{\it Left panel:} schematic representation of the \textsc{RXTorusD} model. The model assumes that the isotropic X-ray source (red star) is surrounded by a torus with a major axis $R$ and a minor axis $r$. The covering factor of this obscurer is given by the ratio between the two axis ($r/R$). The inclination angle $i$ is the angle between the observer and the normal to the plane of the torus. Two column densities can be defined, the line-of-sight column density [$N_{\rm H}(i)$] (see Eq.\,\ref{eq:NHtor}) and the equatorial or maximum column density [$N_{\rm H}(\rm max)=N_{\rm H}(90^{\circ})$]. {\it Right panel:} ratio between the spectra of the \textsc{RXTorusD} model (for \textsc{dust 1}) and those obtained using the same geometry but considering the physical assumption of some of the most-commonly used torus model (i.e., ignoring dust, molecular gas and Rayleigh scattering). All simulations assume an equatorial column density of $N_{\rm H}(\rm max)=4\times10^{24}\rm\,cm^{-2}$, a covering factor of $70\%$ and two different inclination angles: $60-65^{\circ}$ (red continous line) and $85-90^{\circ}$ (black dot-dot-dashed line). The spectral resolution was set to 50\,eV.}
\label{fig:RXTorusDcartoon}
\end{figure*}

With the first version of \reflex\ \citep{Paltani:2017fa} we released a spectral model to reproduce the X-ray emission of AGN, \textsc{RXTorus}, which considered reprocessed radiation produced in a torus with variable covering factor and column density. With this paper we release a new version of the model, \textsc{RXTorusD}, i.e. the first torus model which considers dusty gas. \textsc{RXTorusD} assumes a geometry similar to that of the widely-used \textsc{MYTorus} model, but allowing the covering factor to vary (left panel of Fig.\,\ref{fig:RXTorusDcartoon}), and including more physical processes and more realistic cross-sections (\S\ref{sect:differencesAGNmodels}) with respect to current X-ray models (e.g., \citealp{Murphy:2009ly,Balokovic:2018my,Tanimoto:2019ts}).

\textsc{RXTorusD} was generated by running a grid of simulations with varying parameters, generating $10^{10}$\,photons per simulations. We considered that the X-ray source is pointlike and emits isotropically. The primary X-ray continuum was assumed to have the form of a cutoff power-law, with the photon index $\Gamma$ varying between $1.0$ and $3.0$, while the cutoff energy was fixed to $E_{\rm C}=200$\,keV \citep{Ricci:2018mp}. The maximum energy of the photons was set to be 500\,keV (i.e., no photon was produced above this energy). The inclination angle between the observer and the normal to the plane of the torus ($i$) covers the range $0-90^{\circ}$, with a $3^{\circ}$ interval. Similarly to what was done for \textsc{RXTorus} \citep{Paltani:2017fa}, in order to consider the whole range of inclination angles we assumed that the first and last bin are be centred at 0$^{\circ}$ and 90$^{\circ}$, respectively. The equatorial column density of the torus, i.e. the maximum column density that can be observed for a given configuration [$N_{\rm H}(\mathrm{max})=N_{\rm H}(i=90^{\circ})$], covered the range $10^{22}-10^{25.5}\rm\,cm^{-2}$. The variable covering factor in the model is given by the varying ratio between the minor and major axis ($CF=r/R$), and varies between $0.01$ and $1$. For a given $r/R$ and $i$, the line-of-sight column density is given by
\begin{equation}\label{eq:NHtor}
\left\{\begin{array}{ll}
N_\mathrm{H}(i)=N_\mathrm{H}(\mathrm{max})\left( 1- \frac{R^2}{r^2}  \cos^{2} i \right)^{1/2} , &\cos i<\frac{r}{R},\\
N_\mathrm{H}(i)=0 , &\cos i>\frac{r}{R}\end{array}\right.,
\end{equation}
The gas in the torus was set to be neutral (\textsc{temperature 0} in \reflex), dusty and to have a molecular fraction of $f_{\rm H2}=20\%$ (\textsc{H2fraction 0.2}). We consider two dust depletions: \textsc{dust 0.5} and \textsc{dust 1} (i.e. the case in which iron is fully depleted). We used the proto-solar abundances from \cite{Lodders:2009vs} and assumed three different metallicities (0.3, 1=Solar, 2). The simulations were carried out using a resolution of 20\,eV in the 0.3--8\,keV range. Above 8\,keV, given the lack of narrow spectral features, we used a logarithmic step of $\Delta E/E = 0.001$.

In order to allow for flexibility, besides including a table that contains all simulations, the model is divided into different components: i) the transmitted continuum, containing only photons that did not undergo any interaction with the surrounding medium; ii) the scattered continuum, which includes all photons that underwent scattering, but neither photoionization nor fluorescence; iii) the fluorescence, which contains photons that underwent at least one photoionization and fluorescence event.  The free parameters of the model are the photon index ($\Gamma$) and normalization ($n$) of the primary X-ray emission, the covering factor ($r/R$) and equatorial column density [$N_{\rm H}(\rm max)$] of the torus, and the inclination angle between the observer and the normal to the plane of the torus ($i$).
Following this new release of \reflex, the other \textsc{RXTorus} models have been updated to include the proto-solar abundances of \citet{Lodders:2009vs} and to increase the number of photons per simulation.

\subsubsection{Improvements with respect to commonly-used AGN torus spectral models}\label{sect:differencesAGNmodels}

The main novelty of \textsc{RXTorusD} is that it considers a wide variety of physical processes, which make our simulations extremely realistic. None of the physical X-ray spectral models used to reproduce the emission of AGN currently considers dusty gas. Moreover, some of the most-commonly used models applied to the X-ray spectra of obscured AGN, such as \textsc{MYTorus} \citep{Murphy:2009ly}, \textsc{Borus} \citep{Balokovic:2018my}, \textsc{UxClumpy} \citep{Buchner:2019pf} assume that Compton scattering takes places on free electrons. This assumption is not realistic for cold reprocessing material, such as the circumnuclear dusty medium that surrounds AGN. Our simulations, on the other hand, consider the incoherent scattering functions of \cite{Hubbell1975}, which provide Z- and energy-dependent multiplicative terms to the Klein-Nishina cross-section. The result of these corrections is a reduction of the Compton cross section at low energies, which is more than compensated by the addition of Rayleigh scattering (see \citealp{Paltani:2017fa} for further details). Moreover, our simulations also consider coherent and incoherent scattering on molecular hydrogen. All this can result in large differences in the observed X-ray spectra, particularly below 10\,keV. This is clearly illustrated in the right panel of Fig.\,\ref{fig:RXTorusDcartoon}, which shows the ratio between two spectra from the \textsc{RXTorusD} model and two spectra obtained by considering the same geometrical and physical configuration but ignoring dust, molecular gas and Rayleigh scattering. For the simulations we assumed $N_{\rm H}(\rm max)=4\times10^{24}\rm\,cm^{-2}$, a covering factor of $r/R=0.7$, and considered two inclination angles: $60-65^{\circ}$ (red continous line) and $85-90^{\circ}$ (black dot-dot-dashed line).

\section{Summary and conclusions}

In this work we have reported on the implementation of dust in the ray-tracing platform \reflex\ \citep{Paltani:2017fa}. We have discussed how dusty gas is treated within our simulations, and how it is controlled by a single parameter, \textsc{dust}, which regulates the fraction of metals in dust grains (i.e. the dust depletion). We have discussed the different effects associated to the interaction between X-ray photons and dust grains, such as dust scattering, near-edge X-ray absorption fine structures and shielding (\S\ref{sec:reflex_dust2}), which are all included in \reflex. We have shown how the cross-sections of the photon-gas interaction change depending on the dust depletion factor (\S\ref{sec:crosssectionsdust} and Figs.\,\ref{fig:crosssections},\ref{fig:ratio_crosssections}). Our simulations were tested using some of the most-commonly used X-ray photoelectric absorption models, such as \textsc{phabs} and \textsc{tbabs} (\S\ref{sec:reflexvstbabs} and Fig.\,\ref{fig:phabs_tbabs}).
We have discussed the small effect of dust on the fluorescence lines (\S\ref{sec:fluorescent}, Fig.\,\ref{fig:fluorescentlines} and top panel of Fig.\,\ref{fig:IronKregion}), and showed the imprint left by the presence of dust on the X-ray spectra assuming different geometries (\S\ref{sect:simulations}), which is particularly important below 10\,keV and around the edges (bottom panel of Fig.\,\ref{fig:IronKregion} and Fig.\,\ref{fig:ratiosmodels}). We have shown how \reflex\ can be used to reproduce the X-ray dust scattering halos that are often observed around Galactic sources (\S\ref{sec:dusthalo} and Fig.\,\ref{fig:scatteringhalo}). In the future \reflex\ could be used to create spectral models for X-ray binaries considering different apertures, taking into account dust-scattered radiation.
Finally, we released a new X-ray spectral model (\textsc{RXTorusD}), which is the first torus model that considers dust absorption and scatter, and could be used reproduce the X-ray spectra of obscured active galactic nuclei (\S\ref{sec:RXTorusD}).  \textsc{RXTorusD} considers physical process that are not included in the most widely-used physical models, such as Rayleigh scattering and scattering on molecular gas, which can lead to strong differences in the predicted X-ray spectra for the same set of geometrical and physical parameters (right panel of Fig.\,\ref{fig:RXTorusDcartoon}). The realistic X-ray spectral models that can be created with \reflex\ will allow us to fully exploit the next-generation of X-ray facilities, and in particular with the advent of the high spectral resolution instruments on-board {\it XRISM} (Resolve, \citealp{XRISM-Science-Team:2020fo}) and {\it Athena} (X-IFU, \citealp{Barret:2016ij}).

\begin{acknowledgments}

We thank Bruce Draine for his invaluable insights into the physics of dust. We thank the referee for their insightful feedback, which helped us to improve the quality and clarity of the paper.
CR acknowledges support from the Fondecyt Iniciacion grant 11190831 and ANID BASAL project FB210003. 

\end{acknowledgments}

\bibliography{reflexdust}
\bibliographystyle{aasjournal}

 \end{document}